\documentclass[traditabstract]{aa}
\usepackage{amsmath}
\usepackage{graphicx}
\usepackage[varg]{txfonts}
\usepackage{natbib}
\usepackage{xcolor}
\usepackage[normal]{subfigure}

\usepackage[english]{babel}
\usepackage{epstopdf}
\usepackage[mathscr]{eucal}
\usepackage{amssymb,amsfonts}
\usepackage{txfonts}
\usepackage{float}
\usepackage{color}
\usepackage{url}

\bibpunct{(}{)}{;}{a}{}{,} 
\DeclareMathOperator\arctanh{arctanh}

\newfont{\gwpfont}{cmssq8 scaled 1000}

\begin{document}

\def\aj{AJ}%
\def\araa{ARA\&A}%
\def\apj{ApJ}%
\def\apjl{ApJ}%
\def\apjs{ApJS}%
\def\aap{A\&A}%
 \def\aapr{A\&A~Rev.}%
\def\aaps{A\&AS}%
\def\mnras{MNRAS}
\def\ssr{SSRv}
\def\nat{Nature}
\def\jcap{JCAP}

\def\Mgv{M_{\rm g,500}}
\def\Mg{M_{\rm g}}
\def\YX {Y_{\rm X}}
\def\LXv {L_{\rm X,500}}
\def\TX {T_{\rm X}}
\def\fgv {f_{\rm g,500}}
\def\fg  {f_{\rm g}}
\def\kT {{\rm k}T}
\def\ne {n_{\rm e}}
\def\Mv {M_{\rm 500}}
\def \Rv {R_{500}}
\def\keV {\rm keV}
\def\Yv{Y_{500}}

\def\MT {$M$--$T_{\rm X}$}
\def\MYX {$M$--$Y_{\rm X}$}
\def\MMg {$M_{500}$--$M_{\rm g,500}$}
\def\MgT {$M_{\rm g,500}$--$T_{\rm X}$}
\def\MgY {$M_{\rm g,500}$--$Y_{\rm X}$}

\def\msol {{\rm M_{\odot}}}

\def\lesssim{\mathrel{\hbox{\rlap{\hbox{\lower4pt\hbox{$\sim$}}}\hbox{$<$}}}}
\def\gtrsim{\mathrel{\hbox{\rlap{\hbox{\lower4pt\hbox{$\sim$}}}\hbox{$>$}}}}

\def\psz{MACS J0647.7+7015}

\def\xmm{XMM-{\it Newton}}
\def\planck{{\it Planck}} 
\def\chandra{{\it Chandra}}
\def \rosat {\hbox{\it ROSAT}}
\newcommand{\excpres}{{\gwpfont EXCPRES}}
\newcommand{\ma}[1]{\textcolor{red}{{ #1}}}

\title{Comparison of hydrostatic and lensing cluster mass estimates:\\ a pilot study in MACS J0647.7+7015}

\titlerunning{Revealing systematics in cluster mass estimates}

\author{A.~Ferragamo\inst{1,2,3} \and J.F.~Mac\'{\i}as-P\'{e}rez\inst{4} \thanks{Corresponding author: J.F. Mac\'{\i}as-P\'{e}rez, macias@lpsc.in2p3.fr} \and V.~Pelgrims\inst{4,5,6}  \and F.~Ruppin\inst{3,6} \and M. De~Petris\inst{1} \and F.~Mayet\inst{4} \and M. Mu\~noz-Echeverr\'{\i}a\inst{4} \and L.~Perotto\inst{4} \and E.~Pointecouteau\inst{8}}
\institute{
Dipartimento di Fisica, Sapienza Universit\'a di Roma, Piazzale Aldo Moro 5, I-00185 Roma, Italy 
\and Instituto de Astrof\'{\i}sica de Canarias (IAC), C/ V\'{\i}a L\'actea s/n, E-38205 La Laguna, Tenerife, Spain
  \and Universidad de La Laguna, Departamento de Astrof\'{\i}sica, C/ Astrof\'{\i}sico Francisco S\'anchez s/n, E-38206 La Laguna, Tenerife, Spain
\and Univ. Grenoble Alpes, CNRS, LPSC-IN2P3, 53, avenue des Martyrs, 38000 Grenoble,
France
\and Institute of Astrophysics, Foundation for Research and Technology-Hellas, GR-71110 Heraklion, Greece
\and
Department of Physics, and Institute for Theoretical and Computational Physics, University of Crete, GR-70013 Heraklion, Greece
\and Kavli Institute for Astrophysics and Space Research, Massachusetts Institute of Technology, Cambridge, MA 02139, USA
\and IRAP-Roche,
9, avenue du Colonel Roche BP 44346 31028 Toulouse Cedex 4, France}

\abstract{The detailed characterization of scaling laws relating the observables of cluster of galaxies to
  their mass is crucial for obtaining accurate cosmological constraints with clusters.
  In this paper, we present a comparison between the hydrostatic and lensing mass profiles of the
  cluster \psz\ at $z=0.59$. The hydrostatic mass profile is obtained from the combination of high resolution NIKA2 thermal Sunyaev-Zel'dovich (tSZ) and \xmm\ X-ray observations of the cluster. Instead, the lensing mass profile is obtained from an analysis of the CLASH lensing data based on the lensing convergence map.
  We find significant variation on the cluster mass estimate depending on the observable, the modelling of the data and the knowledge of the cluster dynamical state. This might lead to significant systematic effects on cluster cosmological analyses for which only a single observable is generally used.
  From this pilot study, we conclude that the combination of high resolution SZ, X-ray and lensing data could allow us to identify and correct for these systematic effects. This would constitute a very interesting extension of the NIKA2 SZ Large Program.
  }


\maketitle

\section{Introduction}
\label{sec:introduction}

Clusters of galaxies are the largest gravitationally bound objects in the Universe and constitute the last step of the hierarchical process of structure formation \citep[see][for a review]{2012ARA&A..50..353K}. Therefore, their abundance in mass and redshift and their spatial distribution are powerful cosmological probes. Cluster cosmology is particularly sensitive to the primordial density fluctuations, and the expansion history and matter content of the Universe \citep[see][for a review]{2011ARA&A..49..409A}. Clusters are primarily composed by a dark matter halo, hot baryonic gas and individual galaxies corresponding to about 85, 12 and 3~\% of their total mass, respectively. Thanks to this multi-component nature, the detection and study of clusters of galaxies can be performed via a large number of complementary observables across wavelengths: optical and IR emission from the galaxies in the cluster \citep{1977ARA&A..15..505B}, gravitational lensing effects from radio to optical wavelengths \citep[see][for a review]{2010CQGra..27w3001B}, X-ray emission from the hot baryonic gas \citep{1988xrec.book.....S,2010A&ARv..18..127B}, and the thermal Sunyaev-Zel'dovich (tSZ) effect at microwave and millimeter wavelenghts \citep{SZ1972,SZ1980}.  \\

In the last decade large catalogues of clusters of galaxies have been made available at different wavelengths leading to a large number of cosmological studies \citep[e.g.][]{cluster_counts,2014A&A...571A..21P,2015A&A...574L...8B,2016ApJ...832...95D,2017AJ....153..220B,2018A&A...620A..10P,2019ApJ...878...55B,2019MNRAS.488.4779C,2020PhRvD.102b3509A}. From these studies,
based mainly on cluster number counts as a function of mass and redshift, it can be concluded that cosmological parameter constraints from galaxy clusters may be affected by systematic uncertainties. These uncertainties come both from the theoretical and observational modelling of clusters \citep[see][for a recent summary]{2020arXiv200510204S}.

On the one hand, the modelling of the halo mass function from numerical simulations is not unique and may lead to uncertainties of about 10~\% on the final cosmological parameters. On the other hand, cluster masses are inferred from cluster observables, either directly or via scaling relations. These mass estimates may both be affected by observational and modelling statistical and systematic uncertainties \citep[see][for a review]{2019SSRv..215...25P,Ruppin2019a}.
The Planck 2013 and 2015 results have allowed for a direct comparison of cluster-based and CMB-based cosmology and have shown discrepancy between the two at about $2 \,\sigma$ level \citep{cluster_counts,2014A&A...571A..21P,2016A&A...594A..24P,2016A&A...594A..22P,2018MNRAS.477.4957B}, which has been slightly reduced for the Planck 2018 CMB results \citep{2018arXiv180706205P}. From the cluster analysis point of view, likely explanations for this discrepancy are deviations from the self-similarity and hydrostatic equilibrium (HE) assumptions at the origin of the scaling relations linking the total mass of the cluster to the cluster tSZ emission \citep{2013A&A...557A..52P,plaint13}. For example, one would expect redshift evolution of the scaling laws due to cluster dynamical state and variations on the environmental conditions \citep[e.g.][]{2019A&A...626A..27S}. Such evolution is not generally taken into account to date as most scaling laws have been derived from low redshift clusters \citep[e.g.][]{plaint13}. In particular, cluster dynamical state variations induce significant variations \citep[see][and references therein]{Giulia2020} in the hydrostatic mass bias, which relates the hydrostatic mass to the total mass of the cluster. For the most disturbed clusters the self-similarity approximation may not be valid.  \\

The identification of the dynamical state of clusters requires resolved observations as well as realistic simulations of clusters so that robust indicators can be identified \citep{2020EPJWC.22800008D}. tSZ and X-ray cluster observations can be used to identify inhomogeneities in the ICM, such as over-pressure or high temperature regions as well as shocks. Equivalently, the spatial distribution and velocity of the galaxies in the cluster as well as the  gravitational lensing estimates of the cluster gravitational potential can be used to identify substructures. Moreover, these different observables will lead independent cluster mass estimates with different systematic uncertainties \citep[see][for a detailed description]{2019SSRv..215...25P}.  \\

As a consequence, to extend scaling laws to high redshift clusters there is a need for high resolution observations of a representative sample of clusters at different wavelengths.
This will be the case for the NIKA2 SZ Large program sample that is composed of 45 Planck and ACT SZ-selected clusters in the $0.5 < z < 0.9$ redshift range \citep{2020EPJWC.22800017M,2017ehep.confE..42M}.
NIKA2 is a new generation continuum camera installed at the IRAM 30-m single-dish telescope \citep{ada18, calvo16}. The combination of a large field of view (6.5 arcmin), a high angular resolution (17.7 arcsec at 150 GHz), and a high-sensitivity of $8~\rm{mJy\cdot s^{1/2}}$ at 150~GHz provides the NIKA2 camera with unique tSZ mapping capabilities \citep{Perotto19}.
X-ray high resolution maps of the NIKA2 LP clusters are also being obtained using the \xmm\ satellite. Furthermore, the LPSZ program will be extended with complementary optical and IR data, which are expected to be obtained either from archive data and/or on-purpose observations. Moreover, a NIKA2 LPSZ twin sample composed by synthetic clusters extracted from hydrodynamical simulation dataset are also available \citep{2019A&A...631A..21R}.\\

As a demonstration of the interest of these multi-wavelength observations,
we present in this paper a comparison between the hydrostatic and lensing mass profiles of the
cluster MACS J0647.7+7015, also known as PSZ2 G144.83+25.11.
The former is obtained from the combination of low resolution Planck tSZ, high resolution NIKA2 SZ and \xmm{} X-ray observations of the cluster \citep{rup18}. The latter is obtained from
a re-analysis of the CLASH lensing data \citep{Zitrin15}. 
The paper is organized as follows. Section~\ref{sec:hydromass} presents the different hydrostatic mass estimates used in this paper.
In Section~\ref{sec:lens} we discuss the estimation of the lensing mass profile. The comparison of the different mass profiles estimates is presented in Section~\ref{sec:mass_prof}. Finally, we draw conclusions in Section~\ref{sec:conc}.

\begin{figure}
	\begin{center}
		\includegraphics[width=9cm]{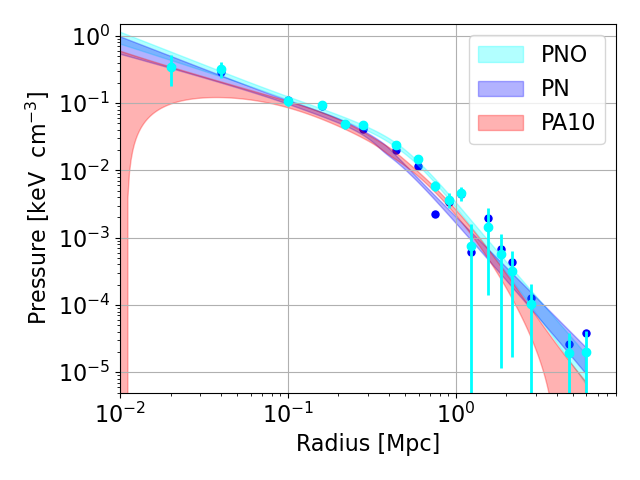}
	\end{center}
	\caption{Pressure profiles derived from the full 150~GHz map (cyan, PNO, cyan) and excluding the over-pressure region (PN, blue). The dots and error bars correspond to the non-parametric fit to the SZ. The shadow region represent the best-fit gNFW model to the non-parametric pressure profiles. We also show in red the universal pressure profile \citep{arn10} for the cluster (PA10).  
  	\label{fig:szp_profiles}}
\end{figure}

\begin{figure*}
	\begin{center}
		\includegraphics[width=18cm]{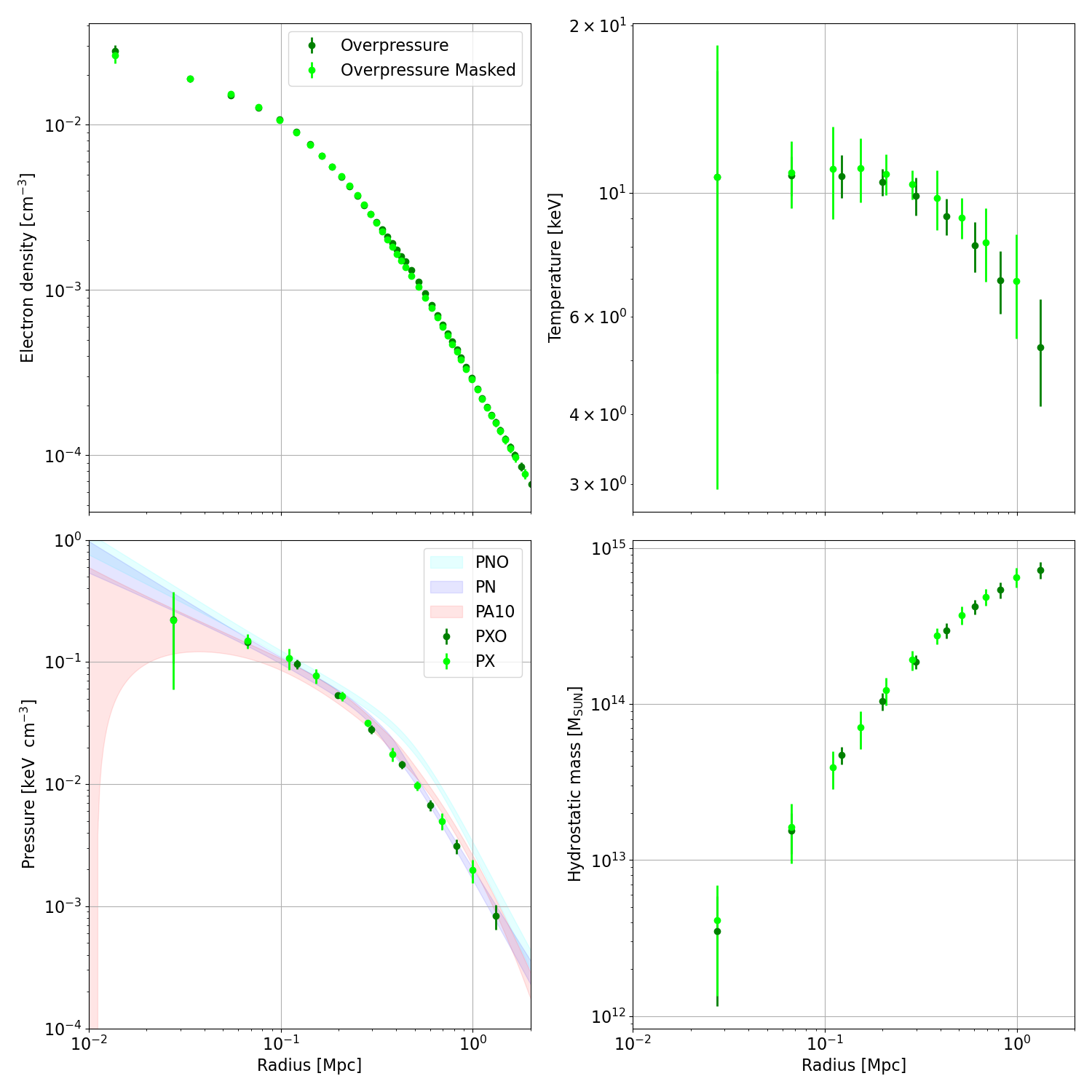}
	\end{center}
	\caption{Electron density, temperature, pressure and hydrostatic mass profiles as derived from the XMM-Newton X-ray data. The dark and light green dots represent the profile computed using the full cluster emission and masking the over-pressure area, respectively. We also show for comparison the PN, PNO and PA10 pressure profiles presented in Fig.~\ref{fig:szp_profiles}.
  	\label{fig:thermo_profiles}}
\end{figure*}

\begin{figure}
	\begin{center}
		\includegraphics[width=9cm]{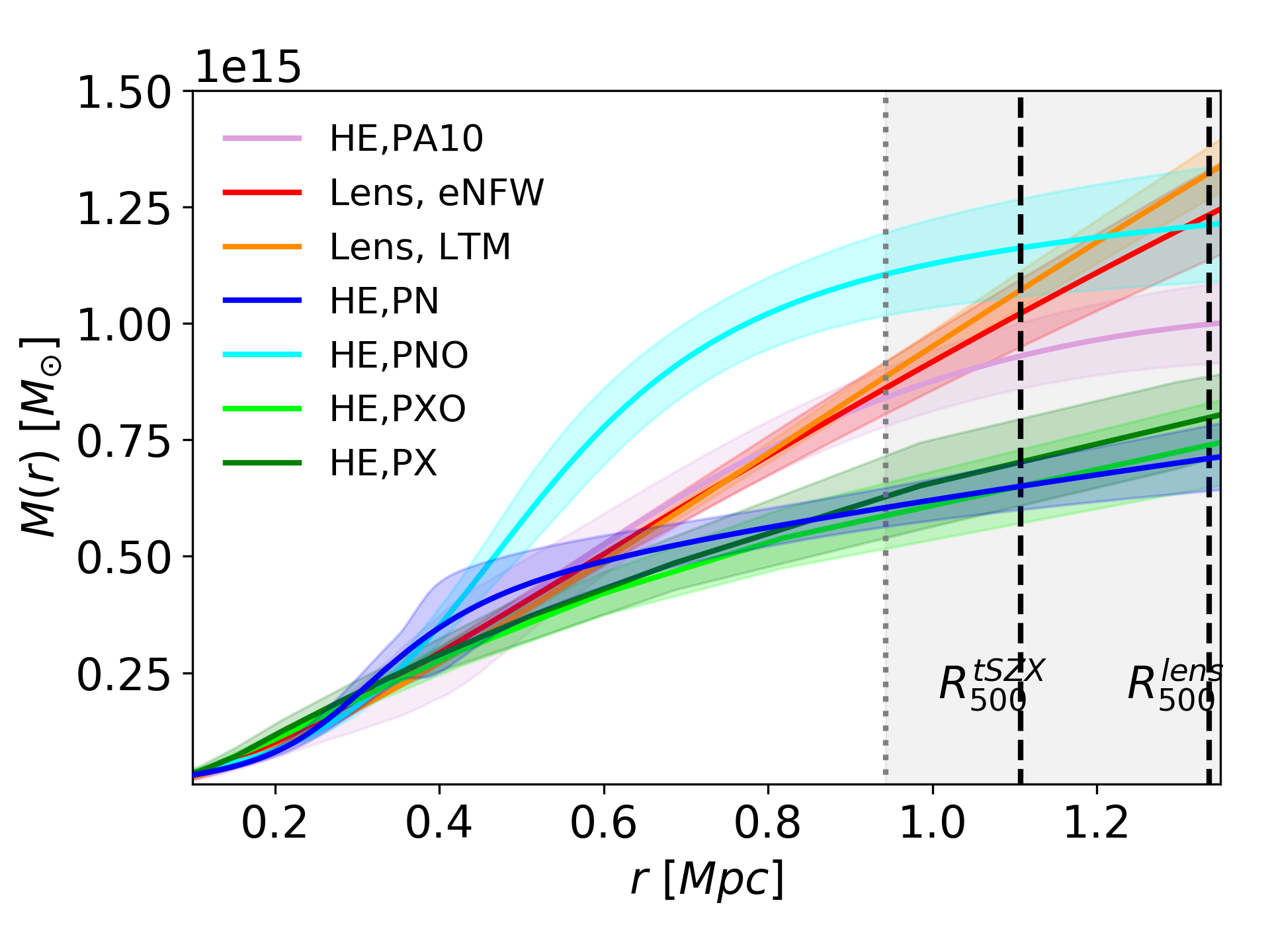}
	\end{center}
	\caption{Combined X-ray and tSZ hydrostatic mass profile computed using three different electron pressure estimates: 1) non-parametric fit using the full 150~GHz NIKA2 map (HE over-pressure, PNO, cyan), 2) non-parametric fit masking the over-pressure region (HE, PN, blue), and, 3) the universal \cite{arn10} pressure profile for this cluster (pink, PA10). Reconstructed X-ray-only hydrostatic mass profiles: excluding (PX, dark green) and including over-pressure (PXO, bright green). Reconstructed lensing mass profile for the eNFW (red) and LTM model (orange). The vertical dotted line represents the maximum radius at which the lensing data are available. The dashed lines show the characteristic radius $R^{\rm{tSZX}}_{\rm{500}}$ and $R^{\rm{lens}}_{\rm{500}}$ (see text for details). 
  	\label{fig:m_profiles}}
\end{figure}

\section{Hydrostatic mass profile}
\label{sec:hydromass}

In this section we discuss and compare different hydrostatic mass radial profiles for the \psz\ cluster.
These were determined from various estimates of the ICM pressure, temperature and electron density radial profiles obtained  
using the Planck and NIKA2 tSZ data, and the \xmm\ X-ray observations as presented in~\citet{rup18}. 
Here, we briefly describe the most important steps of the analysis and present extra material not included in ~\citet{rup18}.

\subsection{tSZ-based pressure profile estimates}\label{sec:Observations}

The tSZ based pressure profile estimates for the ICM of \psz\  have been obtained by \citet{rup18} from a combination of the NIKA2 cluster surface brightness map at 150~GHz \citep{rup18} and the Planck Compton parameter map \citep{2016A&A...594A..22P}. From the NIKA2 brightness map it was found that tSZ signal exhibits an elliptical morphology with a major axis oriented along the E-W direction with a thermal pressure excess in the south-west region. As this over-pressure region might have significant impact in the determination of the hydrostatic mass of the cluster two different thermal pressure profile estimates were considered by excluding or not-excluding the over-pressure region. In both cases the thermal pressure profile were extracted by \citet{rup18} via a Monte Carlo Markov Chain (MCMC) analysis accounting for instrumental properties. Following the work of \cite{rup17}, the deprojected cluster pressure profile was constrained non-parametrically from the cluster core ($\rm{R} \sim 0.02 \rm{R_{500}}$) to its outskirts ($\rm{R} \sim 3 \rm{R_{500}} $). The best fit non-parametric pressure profile was also fitted to a parametric generalized Navarro-Frenk-White model \citep[gNFW, ][]{nag07} to being able to extrapolate at large clustercentric radii. The derived pressure profiles and best-fit gNFW models are presented in Figure~\ref{fig:szp_profiles}. We observe that the pressure profiles are well measured up to 9 Mpc from the center of the cluster, which is well beyond the expected $R_{500}$ radius for the cluster.\\

In summary, we will consider in this work three tSZ-based estimates of the ICM thermal pressure profile of \psz\ : 1) (PNO): NIKA2-Planck combined analysis including the over-pressure region , 2) (PN): NIKA2-Planck combined analysis excluding the over-pressure region , and 3) (PA10): the universal pressure  profile \citep{arn10}.
The latter, which is also presented in Figure~\ref{fig:szp_profiles}, has been obtained by fitting a gNFW model to the data with the slope parameters a, b, and c fixed to the universal pressure  profile values \citep{arn10} and $P_{0}$ and $r_{p}$ as free parameters.

\subsection{X-ray based electron density and pressure profiles}\label{sec:XMM}

Deep X-ray observations of \psz\  by \xmm\ (effective exposure time of ${\sim}68~\rm{ks}$) permit the reconstruction of both the electron density and temperature profiles, from which the pressure profile can be also determined. Here we use the data presented in \citet{rup18} that were processed using the standard procedures presented for example in \citet{bar17}.
We notice that as shown by \citet{rup18} no significant over-density region was identified in the south-west region of \psz\ in the \xmm\ X-ray surface brightness map. However, we will consider here two estimates of the electron density radial profile obtained by excluding or not the over-pressure region.  
The former was already presented in~\citet{rup18}. In both cases the same procedure has been used to constrain the gas density profile from the \xmm\ observations and it is described in detail in \cite{pra10} and \cite{plaint13}. 

The deprojected gas density profiles are fitted by a simplified \citet{vik06} parametric model given by:
\begin{equation}
        n_e(r) = n_{e0} \left[1+\left(\frac{r}{r_c}\right)^2 \right]^{-3 \beta /2} \left[ 1+\left(\frac{r}{r_s}\right)^{\gamma} \right]^{-\epsilon/2 \gamma},
\label{eq:SVM}
\end{equation}
where $n_{e0}$ is the central gas density, $r_c$ is the core radius, and $r_s$ is the transition radius at which an additional steepening in the profile occurs. The $\beta$ and $\epsilon$ parameters define the inner and the outer profile slopes respectively. The $\gamma$ parameter gives the width of the transition in the profile. The value of the $\gamma$ parameter has been fixed to 3 since it provides a good description of all clusters considered in the analysis of \cite{vik06b}.  \\

The \xmm\ data can also be used to estimate the cluster pressure profile by combining the gas density profiles discussed above and the gas temperature determined from the spectroscopic observations via a deprojection procedure as presented in~\cite{pra10} and \cite{plaint13}. In this paper we will use two X-ray based thermal pressure profile estimates for \psz\ : 1) including the over-pressure region (PXO), and 2) excluding the over-pressure region (PX). The latter was already presented in \cite{rup18}. These two pressures profiles allow us to assess the consistency between the X-ray and tSZ views of the ICM.
We show in top row of Figure~\ref{fig:thermo_profiles} the electronic density and temperature as directly derived from the XMM-Newton data set. The dark and light green dots and error bars correspond to the profiles derived from the full cluster data and excluding the over-pressure, respectively. In the bottom row we show the derived pressure profile with the same color scheme. Notice that in this analysis we are limited by the extension of the temperature profile, which can be computed up to about 1.3 (full data) and 1 (excluding over-pressure) Mpc. In the following the hydrostatic mass profile for the X-ray only results will be extrapolated linearly beyond this radius.

\begin{figure*}
 \centering
 \includegraphics[scale=0.65]{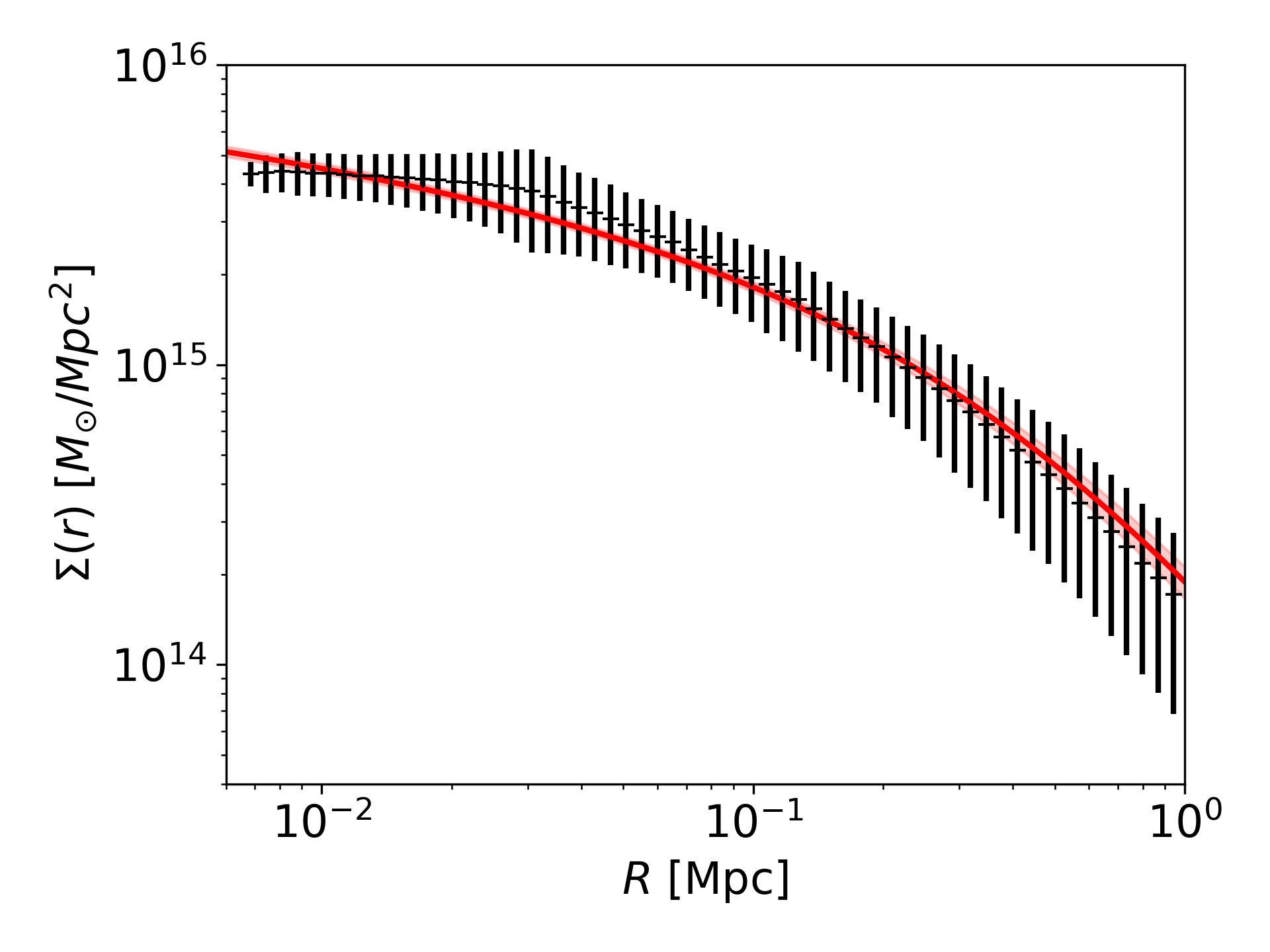}
 \includegraphics[scale=0.55]{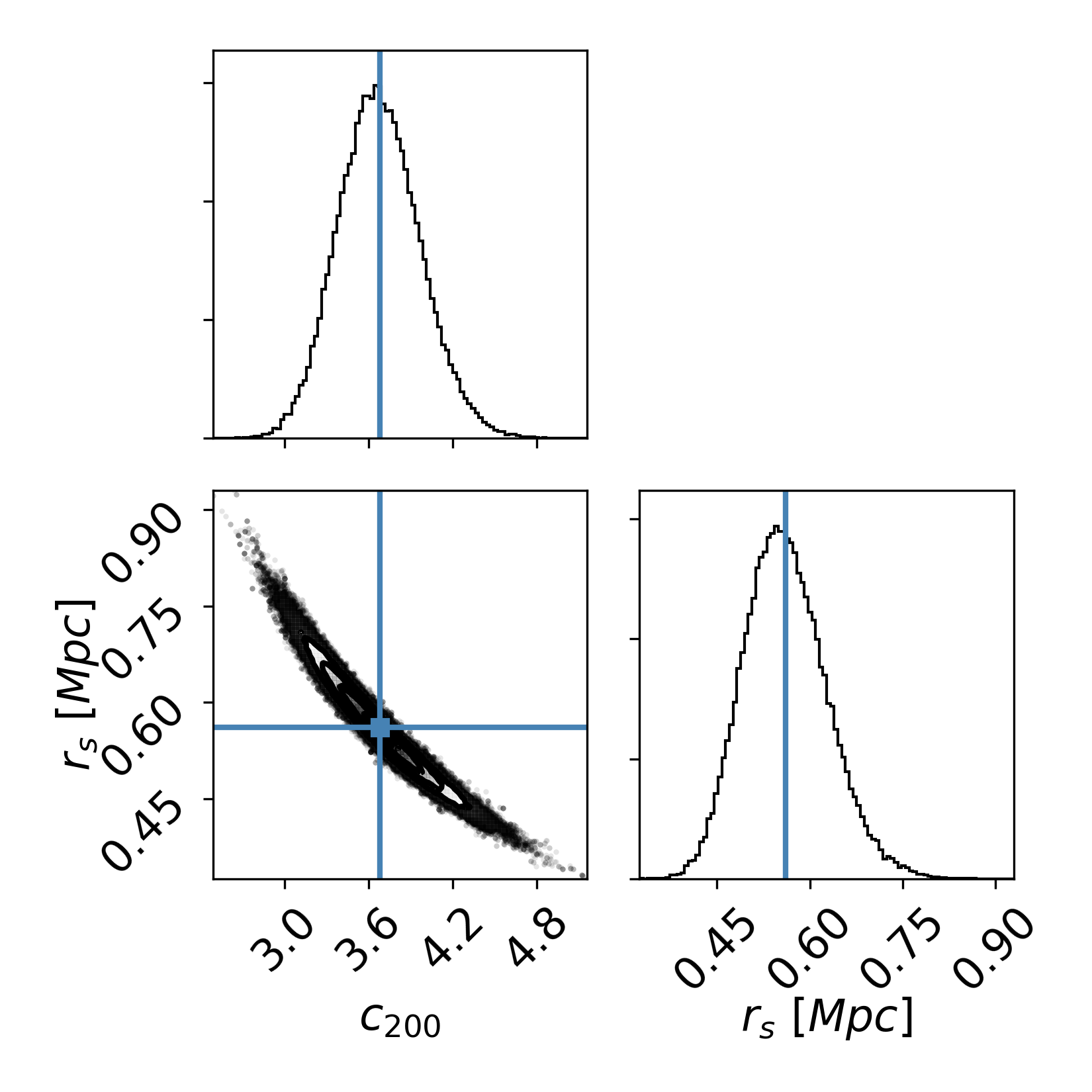}
    \caption{Left: Projected lensing mass density profile as a function of radius, from the {\it PIEMD+eNFW} $\kappa$-map \cite{Zitrin15} lens model (black points). Our best fit projected NFW profile and 1-$\sigma$ uncertainties (red line and shaded area). The 2D sky angular separation to the cluster center has been converted to radial distance in Mpc using the angular distance at cluster redshift. Right: 2D and 1D posterior probability distributions for the $c_{200}$ and $r_{s}$ parameters of the 3 dimensional mass profile as obtained from the MCMC analysis. }
     \label{fig:sigma_prof}
\end{figure*}

\begin{table}
    \centering
    \caption{$R_{500}$ and  $M_{500}$ as derived from the hydrostatic mass profiles for the PN, PNO, PA10, PXO and PX pressure profile estimates.}

    \begin{tabular}{c|c|c}
    \hline
         Pressure profile &  $R^{HE}_{500}$ [kpc] &  $M^{\mathrm{HE}}_{500}$ [ $10^{14}~M_{\sun}$]\\
         \hline
         PN    & $1107 \pm 30$ & $6.95 \pm 0.56$  \\
         PNO   & $1342 \pm 61$ & $12.42 \pm 1.43$ \\
         PA10   & $1236 \pm 178$ & $9.7 \pm 0.8 $ \\
         PXO  & $1070 \pm 40$ & $6.4 \pm 0.8 $ \\
         PX   & $1110 \pm 50$ & $7.0 \pm 0.9 $ \\
         \hline
    \end{tabular}
    \label{tab:hydromassr500}
\end{table}

\subsection{Estimation of the hydrostatic mass profile}\label{sec:HSEmass}

Assuming hydrostatic equilibrium, the total mass enclosed within the radius $r$ is given by:
\begin{equation}
M_{\rm HSE}(r) = -\frac{r^2}{\mu_{\rm{gas}} m_p n_e(r) G} \frac{dP_e(r)}{dr},
\label{eq:hse_mass}
\end{equation}
where $G$ is the gravitational constant, $m_p$ the proton mass, and $\mu_{\rm{gas}} = 0.61$ the mean molecular weight of the gas. 
$n_e$ and $P_e$ are the cluster gas electron density and pressure radial profiles. \\

The hydrostatic mass profile of \psz\, computed using Eq.~\ref{eq:hse_mass}, is shown in Fig~\ref{fig:m_profiles} for the five pressure profiles discussed above: PN, PNO, PA10, PX and PXO. In terms of electron density we have used the electron density profiles discussed above including (in combination with PNO, PA10  and PXO) or excluding (in combination with PN, PX) the over-pressure region. As discussed above the extension of the hydrostatic mass profiles for the X-ray only is limited by the temperature profiles and then they are extrapolated linearly beyond 1.3 and 1 Mpc for the analysis including and excluding the over-pressure region, respectively. To derive the combined X-ray and tSZ hydrostatic mass profiles the non-parametric pressure estimates are previously fitted to a gNFW model (see Figure~\ref{fig:szp_profiles}) and estimates of the pressure derivative are obtained from the best-fit models.

We observe significant differences between the three tSZ and X-ray based estimates (for PN, PNO and PA10) even at large radii, $R_{500}$ and $R_{200}$ at which the cluster masses are traditionally given in the literature. We remind the reader that when including the over-pressure region the tSZ data are significantly not circularly symmetric, thus breaking the hypothesis made for the computation of the pressure and hydrostatic mass profiles. In addition, we find that the two X-ray only based estimates (for PXO and PX) are consistent among them and with the PN based estimate. The over-pressure does not seem to affect the
measured X-ray brightness. \\

We present in Table~\ref{tab:hydromassr500} estimates of the characteristic radius $R^{HE}_{500}$ and of the hydrostatic mass
at this radius, $M^{\mathrm{HE}}_{500}$ for the PN, PNO and PA10 pressure profiles. Notice that the estimates for the PXO and PX
pressure profiles are consistent with those of the PN one. We observe in the table that there are significant differences in the
estimate of the characteristic mass and the radius, which are either related to the over-pressure region or to the differences in the modelling of the pressure profile.
These mass estimates can also be compared to results in the literature based on integrated quantities as $M^{MCXC}_{500}= 8.8 \ 10^{14}~M_{\sun}$  based on $L_{500}$ \citep{2011A&A...534A.109P} and $M^{Planck}_{500}= (8.2 \pm 0.7) \ 10^{14}~M_{\sun}$ based on $Y_{SZ}$ \citep{2016A&A...594A..27P}. 
Although here we present results for a single cluster only, these differences in the mass estimates are illustrative of the systematic uncertainties one would encounter when deriving the hydrostatic mass of a cluster for which we do not dispose of high resolution observations. These uncertainties need to be accounted for when deriving scaling relations relating the cluster integrated observables ($Y_{500}$ and $Y^{X}_{500}$) to the hydrostatic mass. In the following we will define as $\mathrm{R^{tSZX}_{500}}$, the characteristic radius, $R^{HE}_{500}$, for the PN pressure profile estimate and will use it for comparison with the lensing estimate discussed in Sect.\ref{sec:lens}.

\section{Lensing mass density profile}
\label{sec:lens}

In this Section we discuss the reconstruction of the lensing mass density profile of \psz\ using data from the Hubble Space Telescope \citep{CLASH}.
\subsection{Data and preprocessing}

Our analysis is based on the joint Weak and Strong Lensing convergence map (hereafter $\kappa$--map) of \psz\ obtained by \cite{Zitrin15} in their study of the 25 CLASH clusters \citep{CLASH}. 
In \citet{Zitrin15} the authors considered two different parametrizations for the lens model: 1) {\it LTM}, Light-Trace-Mass, and 2) {\it PIEMD+eNFW}, Pseudo Isothermal Elliptical Mass Distribution plus an elliptical Navarro-Frenk-White \citep[NFW;][]{NFW} density profile. More details on these two approaches can be found in \cite {Zitrin09, Zitrin13a, Zitrin13, Zitrin15}. Although we have considered both parametrizations we will limit our presentation here to the analysis to the $\kappa$--map corresponding to the second approach as we find equivalent results. The {\it PIEMD+eNFW} $\kappa$-map of MACS J0647.7+7015 delivered by \citet{Zitrin15} exhibits an elliptical morphology, with a somewhat larger ellipticity than the tSZ NIKA2 map of the cluster, and with the major axis pointing slightly northwards.
\\ \\
The convergence, $\kappa$, is a measure of the projected mass density per unit of critical density at sky position $\vec{\theta}$:
\begin{equation}
\kappa(\vec{\theta}) = \Sigma(\vec{\theta}) / \Sigma_{cr}
\label{eq:kappa}
\end{equation}
with
   \begin{equation}
      \Sigma_{cr} = \frac{c^2}{4 \pi G} \frac{D_s}{D_l D_{ls}},
	\label{eq:sigmacr}
   \end{equation}
where $D_l$, $D_s$ and $D_{ls}$ are respectively the angular diameter distance between the observer and the lens, the observer and the source and between source and lens. $\vec{\theta}$ is the 2D angular separation with respect to the cluster center. \\

Using Eqs.~\ref{eq:kappa} and \ref{eq:sigmacr}, and assuming that the background sources are at redshift of $z_s =2$ \citep{Zitrin11, Zitrin15}, we can directly compute the projected mass density profile, $\Sigma(r)$, from the $\kappa$--map by averaging the signal in rings of increasing distance from the center of the cluster, which is taken to be the same that for the hydrostatic mass analysis. In the left panel of Fig.~\ref{fig:sigma_prof} we show the projected mass density profile obtained by averaging the 
{\it PIEMD+eNFW} $\kappa$ map within 60 rings logarithmically spaced up to a distance from the center of the cluster
of 2.3 arcmin, corresponding to a physical radius $R \sim 1\;Mpc$, (black points). The uncertainties in the projected mass density profile are computed from the standard deviation in each bin in order to account, to first order, for the observed ellipticity in the {\it PIEMD+eNFW} $\kappa$-map.

\subsection{Three-dimensional mass profile reconstruction}

From the projected lensing mass density we can reconstruct the 3 dimensional density profile of the cluster, which we parametrize by using a
3D NFW \citep{NFW} density profile.
Assuming spherical symmetry for the cluster, and using Eqs.~7,~8 and~9 in \cite{Bartelmann96} the projected 3D NFW density profile is given by
	\begin{equation}
		\Sigma(x)=\begin{cases}
		  \frac{2 \rho_s r_s}{x^2-1}\left(1-\frac{2}{\sqrt{x^2-1}}\arctan{\sqrt{\frac{x-1}{x+1}}}\right) & (x>1)\\
		  \frac{2 \rho_s r_s}{x^2-1}\left(1-\frac{2}{\sqrt{1-x^2}}\arctanh{\sqrt{\frac{1-x}{1+x}}}\right) & (x<1)\\
		  \frac{2 \rho_s r_s}{3} & (x=1)
	   \end{cases} 
	\label{eq:proj3DNFW}   
	\end{equation}
as function of the NFW model parameters $\rho_s$ and $r_s$, where $x=r/r_{s}$ and $\rho_s=\delta_c\, \rho_c$. We define
    \begin{equation}
      \delta_c = \frac{200}{3} \frac{c_{\rm{200}}^3}{\ln(1+c_{\rm{200}})-c_{\rm{200}}/(1+c_{\rm{200}})},
   \end{equation}
where $c_{\rm{200}}=r_{\rm{200}}/r_s$ is the concentration parameter and $\rho_c$ the critical density of the Universe.

We fit the numerical projected mass density profile from Fig.~\ref{fig:sigma_prof} to the model in Equation \ref{eq:proj3DNFW} using as free parameters the characteristic cluster radius $r_s$ and the concentration parameter $c_{\rm{200}}$. We perform a Markov Chain Monte Carlo 
analysis (MCMC) to obtain the best-fit parameters. In particular, we used the \texttt{emcee}
MCMC software implemented in Python by \cite{2013ascl.soft03002F} and \cite{Goo2010}. We run the MCMC code using 100 walkers until the convergence
criteria proposed by \cite{Gel1992} is fulfilled for all the model parameters. The 2D and 1D posterior probability distributions for $r_s$ and
$c_{\rm{200}}$ are shown in the right panel of Fig.~\ref{fig:sigma_prof}. The marginalized best-fit values are $r_s=0.56 \pm 0.07\; \rm{Mpc}$ and $c_{\rm{200}}= 3.68 \pm 0.29$. For comparison, we also show in the left panel of Fig.~\ref{fig:sigma_prof} our best-fit model including $1-\sigma$ uncertainties.

Using these best-fit parameters and uncertainties we compute the lensing mass profile of \psz\, which we display as a red line in Fig.~\ref{fig:m_profiles}. Uncertainties at the $1-\sigma$ level are shown as a shaded red area in the figure. For illustration, the corresponding $R^{\mathrm{lens}}_{\rm{500}}= 1340 \pm 52$~kpc is shown as a vertical dashed line. 
We also find $R^{\mathrm{lens}}_{\rm{200}}= 2060 \pm 235$~kpc.
From the mass profile we obtain $M^{\rm{ PIEMD+eNFW,lens}}_{\rm{500}} = (12.3 \pm 1.4)  \times 10^{14}~M_{\sun}$ and the virial mass $M^{\rm{ PIEMD+eNFW,lens}}_{\rm{200}} = (18.1 \pm 2.4) \times 10^{14}~M_{\sun}$. Notice that from the analysis of the {\it LMT} $\kappa$-map  we find slightly larger masses but consistent within uncertainties: $M^{\rm{lens,LMT}}_{\rm{500}} = ( 13.7 \pm 0.9)  \times 10^{14}~M_{\sun}$. Combining the two estimates we obtain  $M^{\rm{lens}}_{\rm{500}} = ( 13.0 \pm  2.8)  \times 10^{14}~M_{\sun}$. We are assuming here that the difference between the two models is a good representation of the modelling systematic uncertainties.
The derived lensing masses are consistent with the one obtained by \citet{Umetsu15}, $M^{\rm{lens}}_{\rm{200}} = (13.2 \pm 4.2)  \, 10^{14}~M_{\sun}$, in their analysis combining strong and weak lensing constraints of a fraction of 20 CLASH clusters where a stacked density profile is fitted.
However our best-fit value of the concentration parameter is more than $2-\sigma$ larger than theirs. Similar conclusions can be drawn from the comparison to the results recently obtained by \cite{2018ApJ...860..104U} for which $M^{\rm{lens}}_{\rm{200}} = ( 11.73 \pm 3.79)  \, 10^{14}~M_{\sun}$ using an {\it eNFW} model. \\ 

We observe that there is good consistency between the different lensing mass estimates. However, we must stress here the fact that numerical simulations indicate that lensing mass estimates may be affected by systematic uncertainties \citep[see][for a review]{2019SSRv..215...25P}. Among those, the uncertainties induced by the mass modelling are probably well represented by the scatter within the different estimates presented here. On the contrary, other like those induced by the uncertainties on the redshift and shape of the cluster galaxies, and on the choice of background galaxies are difficult to estimate for the data used in this paper.

\section{Discussion}
\label{sec:mass_prof}

\subsection{Gas to lensing mass fraction}

\begin{figure}
    \begin{center}
		\includegraphics[width=0.5\textwidth]{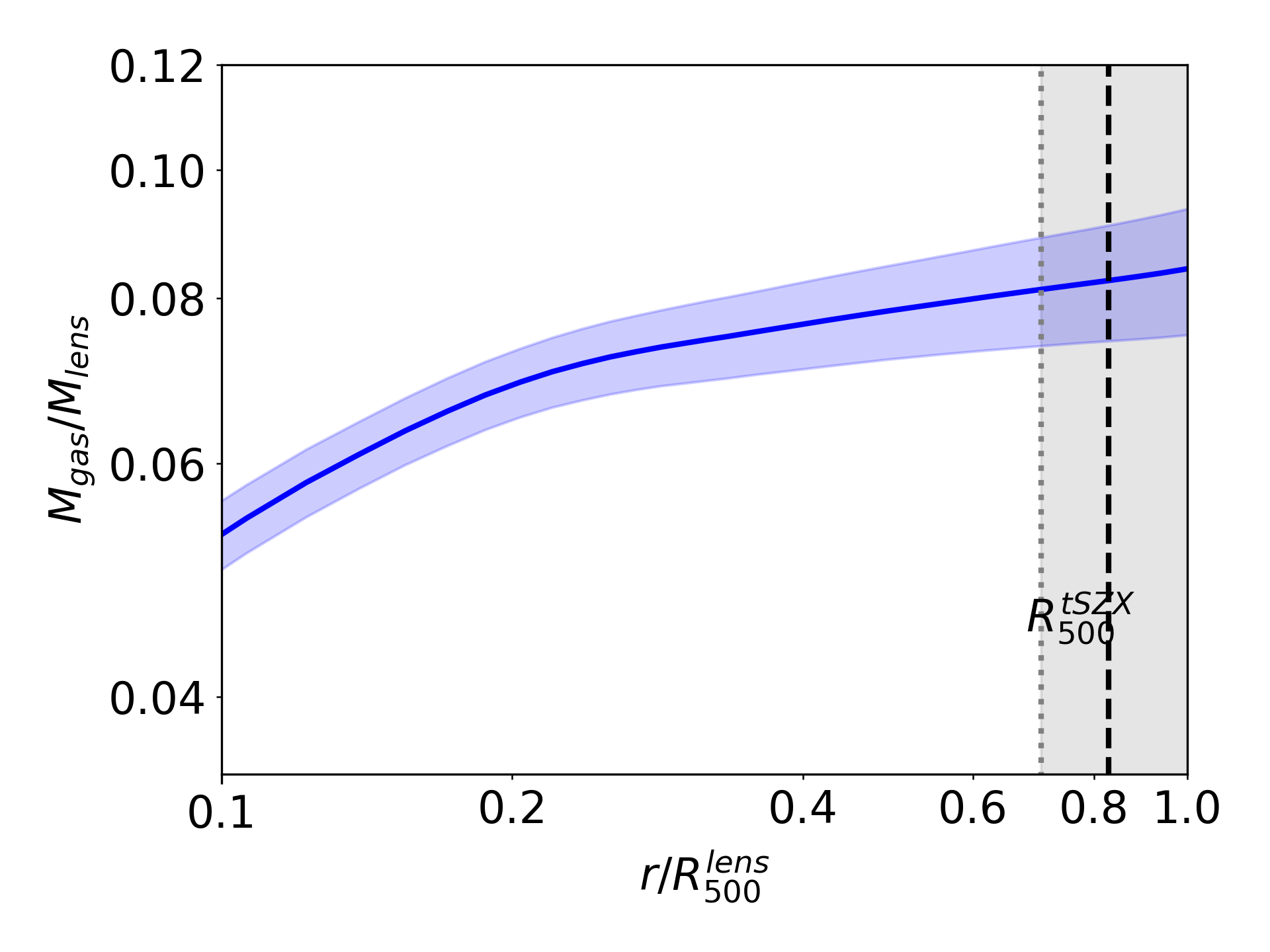}
	\end{center}
	\caption{Gas to lensing mass fraction and uncertainties (blue line and violet shaded area). 
	}
  	\label{fig:fgas}
\end{figure}

From the electron density profile described in Section~\ref{sec:XMM} we can compute the gas mass profile.
Using the lensing mass profile we compute
the gas to lensing mass fraction $\mathrm{f_{gas,lens}}= \frac{M_{\rm{gas}}}{M_{\rm{lens}}} $, which is presented in Figure~\ref{fig:fgas} in blue as function of the normalized radius, assuming $R_{500}^{lens}$ as a reference. 
As both the electron density and the lensing mass profile estimates do not reach large radii we believe that steep increase of the gas fraction observed at those radii is most probably due to a computational artifact.
Uncertainties at the $1-\sigma$ level are shown as a shaded area of the corresponding color.
We find that at $R_{500}^{tSZX}$ the gas fraction is $\mathrm{f_{gas}} = 0.084\pm 0.009$.
This value is compatible with the results of \citet{2016A&A...590L...1P} from a X-ray analysis. It is also marginally in agreement with the results found from the MUSIC hydro-dynamical simulations \citep{2013MNRAS.429..323S} and of \citet{Chiu18} from their analyses of 91 South Pole Telescope selected clusters and with the results of \citet{pressure_profile} on intermediate and low redshift clusters.

Furthermore, by assuming  a stellar fraction of $\sim 6\%$ \citep{Chiu18} and the lensing mass estimate a good proxy for the total mass of the cluster, the resulting baryon fraction is  in agreement with the one form the \citet{2018arXiv180706209P}. 


\subsection{Comparison of the hydrostatic and lensing mass profiles}
In Fig.~\ref{fig:m_profiles} we directly compare the five estimates of the hydrostatic mass profiles and the lensing mass profiles
derived in Sections~\ref{sec:HSEmass}~and~\ref{sec:lens}. We can observe significant differences between these profiles. 
We see that in the case of the joint tSZ and X-ray estimates the over-pressure region increases significantly the hydrostatic mass profile estimate (HE, PNO) as discussed in \citet{rup18}. Such increase results in a larger cluster hydrostatic mass for a given fixed characteristic cluster radius. The estimate of $\mathrm{M_{500}}$ from the lensing analysis is smaller but consistent with the one from the hydrostatic analysis when including the over-pressure (HE, PNO), and nearly a factor of two larger than the one obtained when excluding the over-pressure (HE,PN). They are also a factor of two larger than the ones obtained from the X-ray only estimates both excluding (HE, PX)  and including (HE, PXO) the over-pressure region. \\

In Fig.~\ref{fig:m_profiles} we also show as dashed vertical lines the cluster characteristic radius as computed from the hydrostatic mass profile excluding the over-pressure region (HE, PN), $\mathrm{R_{500}^{tSZX}}$, and the one derived from the lensing mass profile $\mathrm{R_{500}^{lens}}$. We observe that the two estimates are significantly different. \\

The effects discussed above add extra complexity to the cosmological use of the hydrostatic mass estimates. They show
that in practice: 1) the estimates of $\mathrm{R_{500}}$ may vary significantly and as a consequence $\mathrm{M_{500}}$ too, and, 2) the details of the dynamical state of the cluster modify significantly the estimates of the cluster properties and their inter-comparison between different cluster observables. The consequences of the presence of the over-pressure are very different on the hydrostatic mass estimates derived from the tSZ and X-ray joint analysis and from the X-ray only analysis. The study of these systematic effects can not rely on a single cluster and must be
undertaken on a representative sample of clusters. This is one of the major scientific goals of the NIKA2 SZ Large Program \citep{2020EPJWC.22800017M,2017ehep.confE..42M}. \\

\begin{figure}
	\begin{center}
		\includegraphics[scale=0.56]{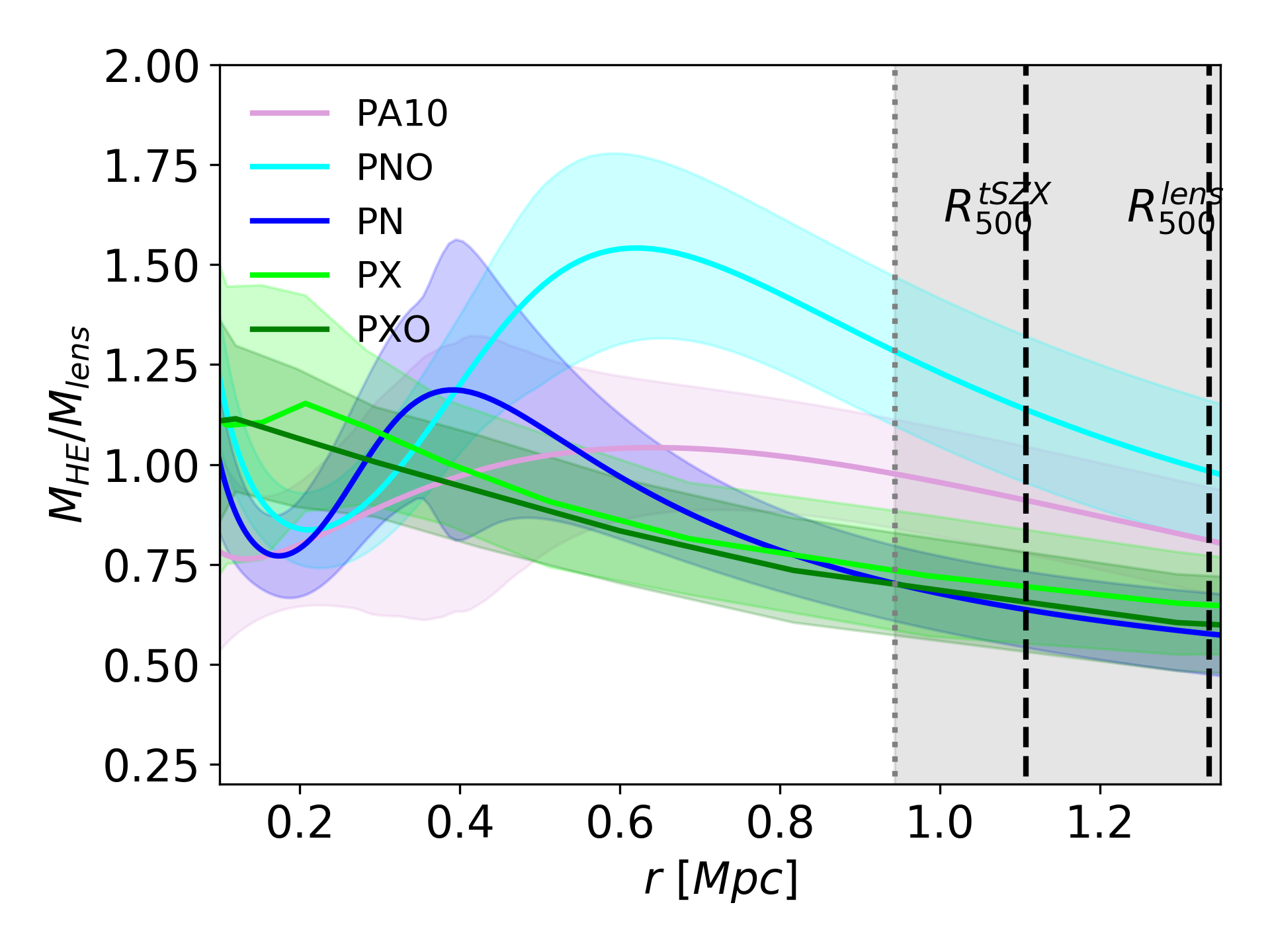}
	\end{center}
	\caption{Hydrostatic to lensing mass ratio for the hydrostatic mass profile estimates for the different pressure profile estimates presented in Sect.~\ref{sec:HSEmass}. Uncertainties at the $1-\sigma$ level are shown as shaded areas of the corresponding color.}
  	\label{fig:fraction_profiles}
\end{figure}

\begin{table}
\caption{Hydrostatic to lensing mass bias $\mathrm{B_{HE-lens}}$ computed from the ratio of the hydrostatic and lensing mass profiles. We compute it for the five hydrostatic mass profile estimates, and, for $\mathrm{R_{500}^{lens}} = 1342 \pm 52$ kpc  and
$\mathrm{R_{500}^{tSZX}} = 1107 \pm 30$ kpc.
}
\begin{tabular}{|c|c c|}
\hline
                   & \multicolumn{2}{|c|}{Hydrostatic to lensing mass bias at} \\
HE mass estimates  &   $\mathrm{R_{500}^{lens}}$  &   $\mathrm{R_{500}^{tSZX}}$ \\
& & \\
\hline
PNO  & $0.98 \pm 0.18$ & $1.14 \pm 0.18$  \\
PN   & $0.58 \pm 0.01$ & $0.64 \pm 0.10$  \\
PA10  & $0.81 \pm 0.13$ & $0.91 \pm 0.13$ \\
PXO & $0.60 \pm 0.12$ & $0.66 \pm 0.12$ \\
PX  & $0.65 \pm 0.12$ & $0.69 \pm 0.14$ \\
\hline
\end{tabular}
\label{tab:hydrolensbias}
\end{table}

\subsection{Hydrostatic to lensing mass bias}
From the results on the gas to lensing mass fraction it seems that the lensing mass might be a good estimate of the total mass of the cluster. Therefore, it is interesting to infer an hydrostatic to lensing mass bias.
In Fig.~\ref{fig:fraction_profiles} we show the ratio between the hydrostatic mass profile and the lensing mass profile, for the five pressure profile estimates discussed in Section~\ref{eq:hse_mass}.
We observe that the $\mathrm{M_{HE}/M_{\rm{lens}}}$ ratio varies from the inner part of the cluster to the outskirt. In particular we find a clear peak for the three hydrostatic mass profile estimates based on the joint tSZ and X-ray analysis, which is probably due to the sharp change of slope that biases towards larger values of the hydrostatic mass in the inner part of the cluster. Significant gradients are also observed in $\mathrm{M_{HE}/M_{lens}}$ ratio between  $\mathrm{R_{500}^{tSZX}}$ and $R_{500}^{lens}$, but it becomes constant at very large radii, typically of the order of $\mathrm{R_{200}}$. We must notice that for these large radii the lensing mass profile is not directly measured but extrapolated from the best-fit model. \\


We compute the hydrostatic to lensing mass bias from the $\mathrm{M_{HE}/M_{lens}}$ ratio for a given characteristic
radius $\mathrm{R_{500}}$. We define it as $\mathrm{B_{HE-lens}= \frac{\mathrm{M^{HE}_{500}}}{\mathrm{M^{lens}_{500}}}} = (1-\mathrm{b_{HE-lens}})$
following the notation in \citet{cluster_counts,2016A&A...594A..22P} for the hydrostatic mass bias.
Building on the previous section discussion we expect this value to vary significantly for the different estimates of the hydrostatic mass. As the estimates of $\mathrm{R_{500}}$ vary significantly, and to further illustrate the possible systematic uncertainties in cosmological analyses, we compute this hydrostatic to lensing mass bias for the two estimates of the characteristic radius $R_{500}$ discussed before: $\mathrm{R_{500}^{tSZX}}$ and $R_{500}^{lens}$. The main results are given in Table~\ref{tab:hydrolensbias}. 
We observe in the table that the hydrostatic to lensing mass bias vary significantly undergoing both positive and negative bias.
These differences are mainly due to the differences in the hydrostatic mass profiles discussed above. However, we also stress the
fact that to compute the hydrostatic to lensing mass bias a common choice of the characteristic radius has to be made and this can also introduce further systematic uncertainties. In Table~\ref{tab:hydrolensbias} we observe that for this cluster the hydrostatic to lensing
mass bias is consistent within uncertainties for the two extreme choices of the characteristic radius $\mathrm{R_{500}}$. \\



\section{Summary and conclusions}
\label{sec:conc}

We have presented in this paper a high resolution multi-wavelength analysis of the cluster of galaxies \psz.
This analysis is a pilot study for the exploitation of the cluster sample from the 
NIKA2 SZ Large program, which will consist on a sample of 45 clusters (24 of which have been already observed) with 
redshift ranging from 0.5 to 0.9 and expected masses, $M_{500}$, from 10$^{14}$ to 10$^{15}$ M$_{\sun}$. \\

We have used high and low resolution tSZ observations from the NIKA2 \citep{rup18} and Planck \citep{2016A&A...594A..22P} experiments, X-ray data from the \xmm\ satellite, and the lensing reconstruction obtained from the Hubble Space Telescope CLASH project \citep{Zitrin15}.
From the projected observables and using MCMC techniques we have been able to derive to high accuracy 
the 3D hydrostatic and lensing mass profiles beyond the $R_{500}$ characteristic radius of the cluster. 
Following the analysis of \cite{rup18} we compute the hydrostatic mass profile including and excluding the
over-pressure region that they identified for different combinations of the tSZ and X-ray data, and for the
universal profile model \citep{arn10}. We find significant differences between the different estimates of the hydrostatic
mass. These differences are related to the assumptions on the dynamical state of the cluster (including or excluding the over-pressure region, the observables used (tSZ and X-ray joint analysis versus X-ray only analysis) and on the pressure profile models considered.\\

From the comparison of hydrostatic and lensing mass profiles of \psz\ we discuss systematic uncertainties
in the estimates of the characteristic quantities representing the cluster like the characteristic radius $R_{500}$, the hydrostatic 
$\mathrm{M^{HE}_{500}}$ and lensing $\mathrm{M^{lens}_{500}}$ masses at that radius, their ratio and the gas fraction. 
We find that the hydrostatic mass estimates $M_{500}$ at the characteristic radius, $R_{500}$, vary by factors of up to two depending on the dynamical state of the cluster (including or excluding the over-pressure region), the choice of the data used (combined tSZ and X-ray versus X-ray only analyses), and the pressure profile estimate.  We also find that the hydrostatic mass estimates are, as expected, overall smaller than those obtained from the lensing mass profile but for the joint tSZ and X-ray analysis when the over-pressure is considered. Equivalently, the estimates of $R_{500}$ vary up to 30 \%. This makes difficult the comparison between different mass estimates given in the literature if no high resolution mass profiles are available. Moreover, these variations can lead to important systematics in the scaling relations used  for cosmological analyses and should be accounted for. \\

We also compute the \psz\ gas to lensing fraction at a fixed characteristic radius, $R_{500}$ and find that it is in a marginally agreement with the results of \citet{pressure_profile} on intermediate and low redshift clusters. However, by assuming that the lensing mass is good proxy of the total mass of the cluster and a stellar fraction $\sim 6\% $ the resulting baryon fraction is compatible with the cosmic one \citep{2018arXiv180706209P}.\\


Finally, we conclude that the above results demonstrate the need for detailed multi-wavelength high resolution observations
of a representative cluster sample at high redshift, as expected from the NIKA2 SZ Large program, in order to fully characterize
the mass-observable scaling relations used in cosmological cluster analyses.


\begin{acknowledgements}
We thanks M. Arnaud, C. Combet and G. Pratt for very useful comments and discussions.
This work has been funded by the European Union’s Horizon 2020 research 
and innovation program under grant agreement number 687312. V.P. also
acknowledges support from the European Research Council under the European
Union’s Horizon 2020 research and innovation program, under grant agreement
No 771282. We also acknowledge funding from the ENIGMASS French LabEx (F.R.).
\end{acknowledgements}

\bibliographystyle{aa} 
\bibliography{mybibliography.bib} 

\begin{thebibliography}{63}
\expandafter\ifx\csname natexlab\endcsname\relax\def\natexlab#1{#1}\fi

\bibitem[{{Abbott} {et~al.}(2020){Abbott}, {Aguena}, {Alarcon}, {Allam},
  {Allen}, {Annis}, {Avila}, {Bacon}, {Bechtol}, {Bermeo}, {Bernstein},
  {Bertin}, {Bhargava}, {Bocquet}, {Brooks}, {Brout}, {Buckley-Geer}, {Burke},
  {Carnero Rosell}, {Carrasco Kind}, {Carretero}, {Castander}, {Cawthon},
  {Chang}, {Chen}, {Choi}, {Costanzi}, {Crocce}, {da Costa}, {Davis}, {De
  Vicente}, {DeRose}, {Desai}, {Diehl}, {Dietrich}, {Dodelson}, {Doel},
  {Drlica-Wagner}, {Eckert}, {Eifler}, {Elvin-Poole}, {Estrada}, {Everett},
  {Evrard}, {Farahi}, {Ferrero}, {Flaugher}, {Fosalba}, {Frieman},
  {Garc{\'\i}a-Bellido}, {Gatti}, {Gaztanaga}, {Gerdes}, {Giannantonio},
  {Giles}, {Grandis}, {Gruen}, {Gruendl}, {Gschwend}, {Gutierrez}, {Hartley},
  {Hinton}, {Hollowood}, {Honscheid}, {Hoyle}, {Huterer}, {James}, {Jarvis},
  {Jeltema}, {Johnson}, {Johnson}, {Kent}, {Krause}, {Kron}, {Kuehn},
  {Kuropatkin}, {Lahav}, {Li}, {Lidman}, {Lima}, {Lin}, {MacCrann}, {Maia},
  {Mantz}, {Marshall}, {Martini}, {Mayers}, {Melchior}, {Mena-Fern{\'a}ndez},
  {Menanteau}, {Miquel}, {Mohr}, {Nichol}, {Nord}, {Ogand o}, {Palmese},
  {Paz-Chinch{\'o}n}, {Plazas}, {Prat}, {Rau}, {Romer}, {Roodman}, {Rooney},
  {Rozo}, {Rykoff}, {Sako}, {Samuroff}, {S{\'a}nchez}, {Sanchez}, {Saro},
  {Scarpine}, {Schubnell}, {Scolnic}, {Serrano}, {Sevilla-Noarbe}, {Sheldon},
  {Smith}, {Smith}, {Suchyta}, {Swanson}, {Tarle}, {Thomas}, {To}, {Troxel},
  {Tucker}, {Varga}, {von der Linden}, {Walker}, {Wechsler}, {Weller},
  {Wilkinson}, {Wu}, {Yanny}, {Zhang}, {Zhang}, {Zuntz}, \& {DES
  Collaboration}}]{2020PhRvD.102b3509A}
{Abbott}, T.~M.~C., {Aguena}, M., {Alarcon}, A., {et~al.} 2020, \prd, 102,
  023509

\bibitem[{{Adam} {et~al.}(2018){Adam}, {Adane}, {Ade}, {Andr{\'e}},
  {Andrianasolo}, {Aussel}, {Beelen}, {Beno{\^i}t}, {Bideaud}, {Billot},
  {Bourrion}, {Bracco}, {Calvo}, {Catalano}, {Coiffard}, {Comis}, {De Petris},
  {D{\'e}sert}, {Doyle}, {Driessen}, {Evans}, {Goupy}, {Kramer}, {Lagache},
  {Leclercq}, {Leggeri}, {Lestrade}, {Mac{\'{\i}}as-P{\'e}rez}, {Mauskopf},
  {Mayet}, {Maury}, {Monfardini}, {Navarro}, {Pascale}, {Perotto}, {Pisano},
  {Ponthieu}, {Rev{\'e}ret}, {Rigby}, {Ritacco}, {Romero}, {Roussel}, {Ruppin},
  {Schuster}, {Sievers}, {Triqueneaux}, {Tucker}, \& {Zylka}}]{ada18}
{Adam}, R., {Adane}, A., {Ade}, P.~A.~R., {et~al.} 2018, \aap, 609, A115

\bibitem[{{Allen} {et~al.}(2011){Allen}, {Evrard}, \&
  {Mantz}}]{2011ARA&A..49..409A}
{Allen}, S.~W., {Evrard}, A.~E., \& {Mantz}, A.~B. 2011, \araa, 49, 409

\bibitem[{{Arnaud} {et~al.}(2010){Arnaud}, {Pratt}, {Piffaretti},
  {B{\"o}hringer}, {Croston}, \& {Pointecouteau}}]{arn10}
{Arnaud}, M., {Pratt}, G.~W., {Piffaretti}, R., {et~al.} 2010, \aap, 517, A92

\bibitem[{{Bahcall}(1977)}]{1977ARA&A..15..505B}
{Bahcall}, N.~A. 1977, \araa, 15, 505

\bibitem[{{Bartalucci} {et~al.}(2017){Bartalucci}, {Arnaud}, {Pratt},
  {Vikhlinin}, {Pointecouteau}, {Forman}, {Jones}, {Mazzotta}, \&
  {Andrade-Santos}}]{bar17}
{Bartalucci}, I., {Arnaud}, M., {Pratt}, G.~W., {et~al.} 2017, \aap, 608, A88

\bibitem[{{Bartelmann}(1996)}]{Bartelmann96}
{Bartelmann}, M. 1996, \aap, 313, 697

\bibitem[{{Bartelmann}(2010)}]{2010CQGra..27w3001B}
{Bartelmann}, M. 2010, Classical and Quantum Gravity, 27, 233001

\bibitem[{{Bocquet} {et~al.}(2019){Bocquet}, {Dietrich}, {Schrabback}, {Bleem},
  {Klein}, {Allen}, {Applegate}, {Ashby}, {Bautz}, {Bayliss}, {Benson},
  {Brodwin}, {Bulbul}, {Canning}, {Capasso}, {Carlstrom}, {Chang}, {Chiu},
  {Cho}, {Clocchiatti}, {Crawford}, {Crites}, {de Haan}, {Desai}, {Dobbs},
  {Foley}, {Forman}, {Garmire}, {George}, {Gladders}, {Gonzalez}, {Grandis},
  {Gupta}, {Halverson}, {Hlavacek-Larrondo}, {Hoekstra}, {Holder}, {Holzapfel},
  {Hou}, {Hrubes}, {Huang}, {Jones}, {Khullar}, {Knox}, {Kraft}, {Lee}, {von
  der Linden}, {Luong-Van}, {Mantz}, {Marrone}, {McDonald}, {McMahon}, {Meyer},
  {Mocanu}, {Mohr}, {Morris}, {Padin}, {Patil}, {Pryke}, {Rapetti},
  {Reichardt}, {Rest}, {Ruhl}, {Saliwanchik}, {Saro}, {Sayre}, {Schaffer},
  {Shirokoff}, {Stalder}, {Stanford}, {Staniszewski}, {Stark}, {Story},
  {Strazzullo}, {Stubbs}, {Vanderlinde}, {Vieira}, {Vikhlinin}, {Williamson},
  \& {Zenteno}}]{2019ApJ...878...55B}
{Bocquet}, S., {Dietrich}, J.~P., {Schrabback}, T., {et~al.} 2019, \apj, 878,
  55

\bibitem[{{B{\"o}hringer} \& {Chon}(2015)}]{2015A&A...574L...8B}
{B{\"o}hringer}, H. \& {Chon}, G. 2015, \aap, 574, L8

\bibitem[{{B{\"o}hringer} {et~al.}(2017){B{\"o}hringer}, {Chon}, {Retzlaff},
  {Tr{\"u}mper}, {Meisenheimer}, \& {Schartel}}]{2017AJ....153..220B}
{B{\"o}hringer}, H., {Chon}, G., {Retzlaff}, J., {et~al.} 2017, \aj, 153, 220

\bibitem[{{B{\"o}hringer} \& {Werner}(2010)}]{2010A&ARv..18..127B}
{B{\"o}hringer}, H. \& {Werner}, N. 2010, \aapr, 18, 127

\bibitem[{{Bolliet} {et~al.}(2018){Bolliet}, {Comis}, {Komatsu}, \&
  {Mac{\'{\i}}as-P{\'e}rez}}]{2018MNRAS.477.4957B}
{Bolliet}, B., {Comis}, B., {Komatsu}, E., \& {Mac{\'{\i}}as-P{\'e}rez}, J.~F.
  2018, \mnras, 477, 4957

\bibitem[{{Calvo} {et~al.}(2016){Calvo}, {Beno{\^\i}t}, {Catalano}, {Goupy},
  {Monfardini}, {Ponthieu}, {Barria}, {Bres}, {Grollier}, {Garde}, {Leggeri},
  {Pont}, {Triqueneaux}, {Adam}, {Bourrion}, {Mac{\'\i}as-P{\'e}rez}, {Rebolo},
  {Ritacco}, {Scordilis}, {Tourres}, {Adane}, {Coiffard}, {Leclercq},
  {D{\'e}sert}, {Doyle}, {Mauskopf}, {Tucker}, {Ade}, {Andr{\'e}}, {Beelen},
  {Belier}, {Bideaud}, {Billot}, {Comis}, {D'Addabbo}, {Kramer}, {Martino},
  {Mayet}, {Pajot}, {Pascale}, {Perotto}, {Rev{\'e}ret}, {Ritacco},
  {Rodriguez}, {Savini}, {Schuster}, {Sievers}, \& {Zylka}}]{calvo16}
{Calvo}, M., {Beno{\^\i}t}, A., {Catalano}, A., {et~al.} 2016, Journal of Low
  Temperature Physics, 184, 816

\bibitem[{{Chiu} {et~al.}(2018){Chiu}, {Mohr}, {McDonald}, {Bocquet}, {Desai},
  {Klein}, {Israel}, {Ashby}, {Stanford}, {Benson}, {Brodwin}, {Abbott},
  {Abdalla}, {Allam}, {Annis}, {Bayliss}, {Benoit-L{\'e}vy}, {Bertin}, {Bleem},
  {Brooks}, {Buckley-Geer}, {Bulbul}, {Capasso}, {Carlstrom}, {Rosell},
  {Carretero}, {Castander}, {Cunha}, {D'Andrea}, {da Costa}, {Davis}, {Diehl},
  {Dietrich}, {Doel}, {Drlica-Wagner}, {Eifler}, {Evrard}, {Flaugher},
  {Garc{\'\i}a-Bellido}, {Garmire}, {Gaztanaga}, {Gerdes}, {Gonzalez}, {Gruen},
  {Gruendl}, {Gschwend}, {Gupta}, {Gutierrez}, {Hlavacek-L}, {Honscheid},
  {James}, {Jeltema}, {Kraft}, {Krause}, {Kuehn}, {Kuhlmann}, {Kuropatkin},
  {Lahav}, {Lima}, {Maia}, {Marshall}, {Melchior}, {Menanteau}, {Miquel},
  {Murray}, {Nord}, {Ogando}, {Plazas}, {Rapetti}, {Reichardt}, {Romer},
  {Roodman}, {Sanchez}, {Saro}, {Scarpine}, {Schindler}, {Schubnell}, {Sharon},
  {Smith}, {Smith}, {Soares-Santos}, {Sobreira}, {Stalder}, {Stern},
  {Strazzullo}, {Suchyta}, {Swanson}, {Tarle}, {Vikram}, {Walker}, {Weller}, \&
  {Zhang}}]{Chiu18}
{Chiu}, I., {Mohr}, J.~J., {McDonald}, M., {et~al.} 2018, \mnras, 478, 3072

\bibitem[{{Costanzi} {et~al.}(2019){Costanzi}, {Rozo}, {Simet}, {Zhang},
  {Evrard}, {Mantz}, {Rykoff}, {Jeltema}, {Gruen}, {Allen}, {McClintock},
  {Romer}, {von der Linden}, {Farahi}, {DeRose}, {Varga}, {Weller}, {Giles},
  {Hollowood}, {Bhargava}, {Bermeo-Hernandez}, {Chen}, {Abbott}, {Abdalla},
  {Avila}, {Bechtol}, {Brooks}, {Buckley-Geer}, {Burke}, {Rosell}, {Kind},
  {Carretero}, {Crocce}, {Cunha}, {da Costa}, {Davis}, {De Vicente}, {Diehl},
  {Dietrich}, {Doel}, {Eifler}, {Estrada}, {Flaugher}, {Fosalba}, {Frieman},
  {Garc{\'\i}a-Bellido}, {Gaztanaga}, {Gerdes}, {Giannantonio}, {Gruendl},
  {Gschwend}, {Gutierrez}, {Hartley}, {Honscheid}, {Hoyle}, {James}, {Krause},
  {Kuehn}, {Kuropatkin}, {Lima}, {Lin}, {Maia}, {March}, {Marshall}, {Martini},
  {Menanteau}, {Miller}, {Miquel}, {Mohr}, {Ogando}, {Plazas}, {Roodman},
  {Sanchez}, {Scarpine}, {Schindler}, {Schubnell}, {Serrano}, {Sevilla-Noarbe},
  {Sheldon}, {Smith}, {Soares-Santos}, {Sobreira}, {Suchyta}, {Swanson},
  {Tarle}, {Thomas}, \& {Wechsler}}]{2019MNRAS.488.4779C}
{Costanzi}, M., {Rozo}, E., {Simet}, M., {et~al.} 2019, \mnras, 488, 4779

\bibitem[{{de Haan} {et~al.}(2016){de Haan}, {Benson}, {Bleem}, {Allen},
  {Applegate}, {Ashby}, {Bautz}, {Bayliss}, {Bocquet}, {Brodwin}, {Carlstrom},
  {Chang}, {Chiu}, {Cho}, {Clocchiatti}, {Crawford}, {Crites}, {Desai},
  {Dietrich}, {Dobbs}, {Doucouliagos}, {Foley}, {Forman}, {Garmire}, {George},
  {Gladders}, {Gonzalez}, {Gupta}, {Halverson}, {Hlavacek-Larrondo},
  {Hoekstra}, {Holder}, {Holzapfel}, {Hou}, {Hrubes}, {Huang}, {Jones},
  {Keisler}, {Knox}, {Lee}, {Leitch}, {von der Linden}, {Luong-Van}, {Mantz},
  {Marrone}, {McDonald}, {McMahon}, {Meyer}, {Mocanu}, {Mohr}, {Murray},
  {Padin}, {Pryke}, {Rapetti}, {Reichardt}, {Rest}, {Ruel}, {Ruhl},
  {Saliwanchik}, {Saro}, {Sayre}, {Schaffer}, {Schrabback}, {Shirokoff},
  {Song}, {Spieler}, {Stalder}, {Stanford}, {Staniszewski}, {Stark}, {Story},
  {Stubbs}, {Vanderlinde}, {Vieira}, {Vikhlinin}, {Williamson}, \&
  {Zenteno}}]{2016ApJ...832...95D}
{de Haan}, T., {Benson}, B.~A., {Bleem}, L.~E., {et~al.} 2016, \apj, 832, 95

\bibitem[{{De Petris} {et~al.}(2020){De Petris}, {Ruppin}, {Sembolini}, {Adam},
  {Baldi}, {Cialone}, {Comis}, {De Luca}, {Gianfagna}, {K{\'e}ruzor{\'e}},
  {Mac{\'\i}as-P{\'e}rez}, {Mayet}, {Perotto}, \&
  {Yepes}}]{2020EPJWC.22800008D}
{De Petris}, M., {Ruppin}, F., {Sembolini}, F., {et~al.} 2020, in European
  Physical Journal Web of Conferences, Vol. 228, European Physical Journal Web
  of Conferences, 00008

\bibitem[{{Foreman-Mackey} {et~al.}(2013){Foreman-Mackey}, {Conley},
  {Meierjurgen Farr}, {Hogg}, {Lang}, {Marshall}, {Price-Whelan}, {Sanders}, \&
  {Zuntz}}]{2013ascl.soft03002F}
{Foreman-Mackey}, D., {Conley}, A., {Meierjurgen Farr}, W., {et~al.} 2013,
  {emcee: The MCMC Hammer}, Astrophysics Source Code Library

\bibitem[{{Gelman} \& {Rubin}(1992)}]{Gel1992}
{Gelman}, A. \& {Rubin}, D.~B. 1992, Statistical Science, 7, 457

\bibitem[{{Gianfagna} {et~al.}(2020){Gianfagna}, {De Petris}, {Yepes}, {De
  Luca}, {Sembolini}, {Cui}, {Biffi}, {K{\'e}ruzor{\'e}},
  {Mac{\'\i}as-P{\'e}rez}, {Mayet}, {Perotto}, {Rasia}, \&
  {Ruppin}}]{Giulia2020}
{Gianfagna}, G., {De Petris}, M., {Yepes}, G., {et~al.} 2020, arXiv e-prints,
  arXiv:2010.03634

\bibitem[{{Goodman} \& {Weare}(2010)}]{Goo2010}
{Goodman}, J. \& {Weare}, J. 2010, Communications in Applied Mathematics and
  Computational Science, Vol.~5, No.~1, p.~65-80, 2010, 5, 65

\bibitem[{{Kravtsov} \& {Borgani}(2012)}]{2012ARA&A..50..353K}
{Kravtsov}, A.~V. \& {Borgani}, S. 2012, \araa, 50, 353

\bibitem[{{Macias-P{\'e}rez} {et~al.}(2017){Macias-P{\'e}rez}, {Adam}, {Ade},
  {Andr{\'e}}, {Arnaud.}, {Aussel}, {Bartalucci}, {Beelen}, {Benoit},
  {Bideaud}, {Bourrion}, {Calvo}, {Catalano}, {Comis}, {D{\'e}sert}, {Doyle},
  {Driessen}, {Goupy}, {Kramer}, {Lagache}, {Leclercq}, {Lestrade}, {Mauskopf},
  {Mayet}, {Monfardini}, {Perotto}, {Pointecouteau}, {Pisano}, {Ponthieu},
  {Pratt}, {Rev{\'e}ret}, {Ritacco}, {Romero}, {Roussel}, {Ruppin}, {Schuster},
  {Sievers}, {Tucker}, {Zylka}, \& {De Petris}}]{2017ehep.confE..42M}
{Macias-P{\'e}rez}, J.~F., {Adam}, R., {Ade}, P., {et~al.} 2017, in Proceedings
  of the European Physical Society Conference on High Energy Physics. 5-12
  July, 2017 Venice, Italy (EPS-HEP2017). Online at <A
  href=``http://pos.sissa.it/cgi-bin/reader/conf.cgi?confid=314''>http://pos.sissa.it/cgi-bin/reader/conf.cgi?confid=314</A>,
  id.42, 42

\bibitem[{{Mayet} {et~al.}(2020){Mayet}, {Adam}, {Ade}, {Andr{\'e}},
  {Andrianasolo}, {Arnaud}, {Aussel}, {Bartalucci}, {Beelen}, {Beno{\^\i}t},
  {Bideaud}, {Bourrion}, {Calvo}, {Catalano}, {Comis}, {De Petris},
  {D{\'e}sert}, {Doyle}, {Driessen}, {Gomez}, {Goupy}, {K{\'e}ruzor{\'e}},
  {Kramer}, {Ladjelate}, {Lagache}, {Leclercq}, {Lestrade},
  {Mac{\'\i}as-P{\'e}rez}, {Mauskopf}, {Monfardini}, {Perotto}, {Pisano},
  {Pointecouteau}, {Ponthieu}, {Pratt}, {Rev{\'e}ret}, {Ritacco}, {Romero},
  {Roussel}, {Ruppin}, {Schuster}, {Shu}, {Sievers}, {Tucker}, \&
  {Zylka}}]{2020EPJWC.22800017M}
{Mayet}, F., {Adam}, R., {Ade}, P., {et~al.} 2020, in European Physical Journal
  Web of Conferences, Vol. 228, European Physical Journal Web of Conferences,
  00017

\bibitem[{{Nagai} {et~al.}(2007){Nagai}, {Kravtsov}, \& {Vikhlinin}}]{nag07}
{Nagai}, D., {Kravtsov}, A.~V., \& {Vikhlinin}, A. 2007, \apj, 668, 1

\bibitem[{{Navarro} {et~al.}(1996){Navarro}, {Frenk}, \& {White}}]{NFW}
{Navarro}, J.~F., {Frenk}, C.~S., \& {White}, S.~D.~M. 1996, \apj, 462, 563

\bibitem[{{Pacaud} {et~al.}(2018){Pacaud}, {Pierre}, {Melin}, {Adami},
  {Evrard}, {Galli}, {Gastaldello}, {Maughan}, {Sereno}, {Alis}, {Altieri},
  {Birkinshaw}, {Chiappetti}, {Faccioli}, {Giles}, {Horellou}, {Iovino},
  {Koulouridis}, {Le F{\`e}vre}, {Lidman}, {Lieu}, {Maurogordato},
  {Moscardini}, {Plionis}, {Poggianti}, {Pompei}, {Sadibekova}, {Valtchanov},
  \& {Willis}}]{2018A&A...620A..10P}
{Pacaud}, F., {Pierre}, M., {Melin}, J.~B., {et~al.} 2018, \aap, 620, A10

\bibitem[{{Perotto} {et~al.}(2020){Perotto}, {Ponthieu},
  {Mac{\'\i}as-P{\'e}rez}, {Adam}, {Ade}, {Andr{\'e}}, {Andrianasolo},
  {Aussel}, {Beelen}, {Beno{\^\i}t}, {Berta}, {Bideaud}, {Bourrion}, {Calvo},
  {Catalano}, {Comis}, {De Petris}, {D{\'e}sert}, {Doyle}, {Driessen},
  {Garc{\'\i}a}, {Gomez}, {Goupy}, {John}, {K{\'e}ruzor{\'e}}, {Kramer},
  {Ladjelate}, {Lagache}, {Leclercq}, {Lestrade}, {Maury}, {Mauskopf}, {Mayet},
  {Monfardini}, {Navarro}, {Pe{\~n}alver}, {Pierfederici}, {Pisano},
  {Rev{\'e}ret}, {Ritacco}, {Romero}, {Roussel}, {Ruppin}, {Schuster}, {Shu},
  {Sievers}, {Tucker}, \& {Zylka}}]{Perotto19}
{Perotto}, L., {Ponthieu}, N., {Mac{\'\i}as-P{\'e}rez}, J.~F., {et~al.} 2020,
  \aap, 637, A71

\bibitem[{{Piffaretti} {et~al.}(2011){Piffaretti}, {Arnaud}, {Pratt},
  {Pointecouteau}, \& {Melin}}]{2011A&A...534A.109P}
{Piffaretti}, R., {Arnaud}, M., {Pratt}, G.~W., {Pointecouteau}, E., \&
  {Melin}, J.~B. 2011, \aap, 534, A109

\bibitem[{{Planck Collaboration} {et~al.}(2014{\natexlab{a}}){Planck
  Collaboration}, {Ade}, {Aghanim}, {Armitage-Caplan}, {Arnaud}, {Ashdown},
  {Atrio-Barandela}, {Aumont}, {Baccigalupi}, {Banday}, \&
  et~al.}]{cluster_counts}
{Planck Collaboration}, {Ade}, P.~A.~R., {Aghanim}, N., {et~al.}
  2014{\natexlab{a}}, \aap, 571, A20

\bibitem[{{Planck Collaboration} {et~al.}(2014{\natexlab{b}}){Planck
  Collaboration}, {Ade}, {Aghanim}, {Armitage-Caplan}, {Arnaud}, {Ashdown},
  {Atrio-Barandela}, {Aumont}, {Baccigalupi}, {Banday}, \&
  et~al.}]{2014A&A...571A..21P}
{Planck Collaboration}, {Ade}, P.~A.~R., {Aghanim}, N., {et~al.}
  2014{\natexlab{b}}, \aap, 571, A21

\bibitem[{{Planck Collaboration} {et~al.}(2013{\natexlab{a}}){Planck
  Collaboration}, {Ade}, {Aghanim}, {Arnaud}, {Ashdown}, {Atrio-Barandela},
  {Aumont}, {Baccigalupi}, {Balbi}, {Banday}, {Barreiro}, {Bartlett},
  {Battaner}, {Battye}, {Benabed}, {Bernard}, {Bersanelli}, {Bhatia},
  {Bikmaev}, {B{\"o}hringer}, {Bonaldi}, {Bond}, {Borgani}, {Borrill},
  {Bouchet}, {Bourdin}, {Brown}, {Bucher}, {Burenin}, {Burigana}, {Butler},
  {Cabella}, {Cardoso}, {Carvalho}, {Chamballu}, {Chiang}, {Chon}, {Clements},
  {Colafrancesco}, {Coulais}, {Cuttaia}, {Da Silva}, {Dahle}, {Davis}, {de
  Bernardis}, {de Gasperis}, {Delabrouille}, {D{\'e}mocl{\`e}s}, {D{\'e}sert},
  {Diego}, {Dolag}, {Dole}, {Donzelli}, {Dor{\'e}}, {Douspis}, {Dupac},
  {Efstathiou}, {En{\ss}lin}, {Eriksen}, {Finelli}, {Flores-Cacho}, {Forni},
  {Frailis}, {Franceschi}, {Frommert}, {Galeotta}, {Ganga},
  {G{\'e}nova-Santos}, {Giard}, {Giraud-H{\'e}raud}, {Gonz{\'a}lez-Nuevo},
  {G{\'o}rski}, {Gregorio}, {Gruppuso}, {Hansen}, {Harrison},
  {Hern{\'a}ndez-Monteagudo}, {Herranz}, {Hildebrandt}, {Hivon}, {Hobson},
  {Holmes}, {Huffenberger}, {Hurier}, {Jagemann}, {Juvela}, {Keih{\"a}nen},
  {Khamitov}, {Kneissl}, {Knoche}, {Kunz}, {Kurki-Suonio}, {Lagache},
  {Lamarre}, {Lasenby}, {Lawrence}, {Le Jeune}, {Leach}, {Leonardi}, {Liddle},
  {Lilje}, {Linden-V{\o}rnle}, {L{\'o}pez-Caniego}, {Luzzi},
  {Mac{\'{\i}}as-P{\'e}rez}, {Maino}, {Mandolesi}, {Maris}, {Marleau},
  {Marshall}, {Mart{\'{\i}}nez-Gonz{\'a}lez}, {Masi}, {Matarrese}, {Matthai},
  {Mazzotta}, {Meinhold}, {Melchiorri}, {Melin}, {Mendes}, {Mitra},
  {Miville-Desch{\^e}nes}, {Montier}, {Morgante}, {Munshi}, {Natoli},
  {N{\o}rgaard-Nielsen}, {Noviello}, {Osborne}, {Pajot}, {Paoletti},
  {Partridge}, {Pearson}, {Perdereau}, {Perrotta}, {Piacentini}, {Piat},
  {Pierpaoli}, {Piffaretti}, {Platania}, {Pointecouteau}, {Polenta},
  {Ponthieu}, {Popa}, {Poutanen}, {Pratt}, {Prunet}, {Puget}, {Rachen},
  {Rebolo}, {Reinecke}, {Remazeilles}, {Renault}, {Ricciardi}, {Ristorcelli},
  {Rocha}, {Rosset}, {Rossetti}, {Rubi{\~n}o-Mart{\'{\i}}n}, {Rusholme},
  {Sandri}, {Savini}, {Scott}, {Starck}, {Stivoli}, {Stolyarov}, {Sudiwala},
  {Sunyaev}, {Sutton}, {Suur-Uski}, {Sygnet}, {Tauber}, {Terenzi},
  {Toffolatti}, {Tomasi}, {Tristram}, {Valenziano}, {Van Tent}, {Vielva},
  {Villa}, {Vittorio}, {Wandelt}, {Weller}, {White}, {Yvon}, {Zacchei}, \&
  {Zonca}}]{plaint13}
{Planck Collaboration}, {Ade}, P.~A.~R., {Aghanim}, N., {et~al.}
  2013{\natexlab{a}}, \aap, 550, A129

\bibitem[{{Planck Collaboration} {et~al.}(2013{\natexlab{b}}){Planck
  Collaboration}, {Ade}, {Aghanim}, {Arnaud}, {Ashdown}, {Atrio-Barandela},
  {Aumont}, {Baccigalupi}, {Balbi}, {Banday}, \& et~al.}]{pressure_profile}
{Planck Collaboration}, {Ade}, P.~A.~R., {Aghanim}, N., {et~al.}
  2013{\natexlab{b}}, \aap, 550, A131

\bibitem[{{Planck Collaboration} {et~al.}(2013{\natexlab{c}}){Planck
  Collaboration}, {Ade}, {Aghanim}, {Arnaud}, {Ashdown}, {Atrio-Barandela},
  {Aumont}, {Baccigalupi}, {Balbi}, {Banday}, \& et~al.}]{2013A&A...557A..52P}
{Planck Collaboration}, {Ade}, P.~A.~R., {Aghanim}, N., {et~al.}
  2013{\natexlab{c}}, \aap, 557, A52

\bibitem[{{Planck Collaboration} {et~al.}(2016{\natexlab{a}}){Planck
  Collaboration}, {Ade}, {Aghanim}, {Arnaud}, {Ashdown}, {Aumont},
  {Baccigalupi}, {Banday}, {Barreiro}, {Barrena}, \&
  et~al.}]{2016A&A...594A..27P}
{Planck Collaboration}, {Ade}, P.~A.~R., {Aghanim}, N., {et~al.}
  2016{\natexlab{a}}, \aap, 594, A27

\bibitem[{{Planck Collaboration} {et~al.}(2016{\natexlab{b}}){Planck
  Collaboration}, {Ade}, {Aghanim}, {Arnaud}, {Ashdown}, {Aumont},
  {Baccigalupi}, {Banday}, {Barreiro}, {Bartlett}, \&
  et~al.}]{2016A&A...594A..24P}
{Planck Collaboration}, {Ade}, P.~A.~R., {Aghanim}, N., {et~al.}
  2016{\natexlab{b}}, \aap, 594, A24

\bibitem[{{Planck Collaboration} {et~al.}(2018{\natexlab{a}}){Planck
  Collaboration}, {Aghanim}, {Akrami}, {Ashdown}, {Aumont}, {Baccigalupi},
  {Ballardini}, {Banday}, {Barreiro}, {Bartolo}, {Basak}, {Battye}, {Benabed},
  {Bernard}, {Bersanelli}, {Bielewicz}, {Bock}, {Bond}, {Borrill}, {Bouchet},
  {Boulanger}, {Bucher}, {Burigana}, {Butler}, {Calabrese}, {Cardoso},
  {Carron}, {Challinor}, {Chiang}, {Chluba}, {Colombo}, {Combet}, {Contreras},
  {Crill}, {Cuttaia}, {de Bernardis}, {de Zotti}, {Delabrouille}, {Delouis},
  {Di Valentino}, {Diego}, {Dor{\'e}}, {Douspis}, {Ducout}, {Dupac}, {Dusini},
  {Efstathiou}, {Elsner}, {En{\ss}lin}, {Eriksen}, {Fantaye}, {Farhang},
  {Fergusson}, {Fernandez-Cobos}, {Finelli}, {Forastieri}, {Frailis},
  {Franceschi}, {Frolov}, {Galeotta}, {Ganga}, {G{\'e}nova-Santos}, {Gerbino},
  {Ghosh}, {Gonz{\'a}lez-Nuevo}, {G{\'o}rski}, {Gratton}, {Gruppuso},
  {Gudmundsson}, {Hamann}, {Hand ley}, {Herranz}, {Hivon}, {Huang}, {Jaffe},
  {Jones}, {Karakci}, {Keih{\"a}nen}, {Keskitalo}, {Kiiveri}, {Kim}, {Kisner},
  {Knox}, {Krachmalnicoff}, {Kunz}, {Kurki-Suonio}, {Lagache}, {Lamarre},
  {Lasenby}, {Lattanzi}, {Lawrence}, {Le Jeune}, {Lemos}, {Lesgourgues},
  {Levrier}, {Lewis}, {Liguori}, {Lilje}, {Lilley}, {Lindholm},
  {L{\'o}pez-Caniego}, {Lubin}, {Ma}, {Mac{\'\i}as-P{\'e}rez}, {Maggio},
  {Maino}, {Mandolesi}, {Mangilli}, {Marcos-Caballero}, {Maris}, {Martin},
  {Martinelli}, {Mart{\'\i}nez-Gonz{\'a}lez}, {Matarrese}, {Mauri}, {McEwen},
  {Meinhold}, {Melchiorri}, {Mennella}, {Migliaccio}, {Millea}, {Mitra},
  {Miville-Desch{\^e}nes}, {Molinari}, {Montier}, {Morgante}, {Moss}, {Natoli},
  {N{\o}rgaard-Nielsen}, {Pagano}, {Paoletti}, {Partridge}, {Patanchon},
  {Peiris}, {Perrotta}, {Pettorino}, {Piacentini}, {Polastri}, {Polenta},
  {Puget}, {Rachen}, {Reinecke}, {Remazeilles}, {Renzi}, {Rocha}, {Rosset},
  {Roudier}, {Rubi{\~n}o-Mart{\'\i}n}, {Ruiz-Granados}, {Salvati}, {Sandri},
  {Savelainen}, {Scott}, {Shellard}, {Sirignano}, {Sirri}, {Spencer},
  {Sunyaev}, {Suur-Uski}, {Tauber}, {Tavagnacco}, {Tenti}, {Toffolatti},
  {Tomasi}, {Trombetti}, {Valenziano}, {Valiviita}, {Van Tent}, {Vibert},
  {Vielva}, {Villa}, {Vittorio}, {Wand elt}, {Wehus}, {White}, {White},
  {Zacchei}, \& {Zonca}}]{2018arXiv180706209P}
{Planck Collaboration}, {Aghanim}, N., {Akrami}, Y., {et~al.}
  2018{\natexlab{a}}, arXiv e-prints, arXiv:1807.06209

\bibitem[{{Planck Collaboration} {et~al.}(2016{\natexlab{c}}){Planck
  Collaboration}, {Aghanim}, {Arnaud}, {Ashdown}, {Aumont}, {Baccigalupi},
  {Banday}, {Barreiro}, {Bartlett}, {Bartolo}, \& et~al.}]{2016A&A...594A..22P}
{Planck Collaboration}, {Aghanim}, N., {Arnaud}, M., {et~al.}
  2016{\natexlab{c}}, \aap, 594, A22

\bibitem[{{Planck Collaboration} {et~al.}(2018{\natexlab{b}}){Planck
  Collaboration}, {Akrami}, {Arroja}, {Ashdown}, {Aumont}, {Baccigalupi},
  {Ballardini}, {Banday}, {Barreiro}, {Bartolo}, {Basak}, {Battye}, {Benabed},
  {Bernard}, {Bersanelli}, {Bielewicz}, {Bock}, {Bond}, {Borrill}, {Bouchet},
  {Boulanger}, {Bucher}, {Burigana}, {Butler}, {Calabrese}, {Cardoso},
  {Carron}, {Casaponsa}, {Challinor}, {Chiang}, {Colombo}, {Combet},
  {Contreras}, {Crill}, {Cuttaia}, {de Bernardis}, {de Zotti}, {Delabrouille},
  {Delouis}, {D{\'e}sert}, {Di Valentino}, {Dickinson}, {Diego}, {Donzelli},
  {Dor{\'e}}, {Douspis}, {Ducout}, {Dupac}, {Efstathiou}, {Elsner},
  {En{\ss}lin}, {Eriksen}, {Falgarone}, {Fantaye}, {Fergusson},
  {Fernandez-Cobos}, {Finelli}, {Forastieri}, {Frailis}, {Franceschi},
  {Frolov}, {Galeotta}, {Galli}, {Ganga}, {G{\'e}nova-Santos}, {Gerbino},
  {Ghosh}, {Gonz{\'a}lez-Nuevo}, {G{\'o}rski}, {Gratton}, {Gruppuso},
  {Gudmundsson}, {Hamann}, {Hand ley}, {Hansen}, {Helou}, {Herranz}, {Hivon},
  {Huang}, {Jaffe}, {Jones}, {Karakci}, {Keih{\"a}nen}, {Keskitalo}, {Kiiveri},
  {Kim}, {Kisner}, {Knox}, {Krachmalnicoff}, {Kunz}, {Kurki-Suonio}, {Lagache},
  {Lamarre}, {Langer}, {Lasenby}, {Lattanzi}, {Lawrence}, {Le Jeune}, {Leahy},
  {Lesgourgues}, {Levrier}, {Lewis}, {Liguori}, {Lilje}, {Lilley}, {Lindholm},
  {L{\'o}pez-Caniego}, {Lubin}, {Ma}, {Mac{\'\i}as-P{\'e}rez}, {Maggio},
  {Maino}, {Mand olesi}, {Mangilli}, {Marcos-Caballero}, {Maris}, {Martin},
  {Mart{\'\i}nez-Gonz{\'a}lez}, {Matarrese}, {Mauri}, {McEwen}, {Meerburg},
  {Meinhold}, {Melchiorri}, {Mennella}, {Migliaccio}, {Millea}, {Mitra},
  {Miville-Desch{\^e}nes}, {Molinari}, {Moneti}, {Montier}, {Morgante}, {Moss},
  {Mottet}, {M{\"u}nchmeyer}, {Natoli}, {N{\o}rgaard-Nielsen}, {Oxborrow},
  {Pagano}, {Paoletti}, {Partridge}, {Patanchon}, {Pearson}, {Peel}, {Peiris},
  {Perrotta}, {Pettorino}, {Piacentini}, {Polastri}, {Polenta}, {Puget},
  {Rachen}, {Reinecke}, {Remazeilles}, {Renzi}, {Rocha}, {Rosset}, {Roudier},
  {Rubi{\~n}o-Mart{\'\i}n}, {Ruiz-Granados}, {Salvati}, {Sandri}, {Savelainen},
  {Scott}, {Shellard}, {Shiraishi}, {Sirignano}, {Sirri}, {Spencer}, {Sunyaev},
  {Suur-Uski}, {Tauber}, {Tavagnacco}, {Tenti}, {Terenzi}, {Toffolatti},
  {Tomasi}, {Trombetti}, {Valiviita}, {Van Tent}, {Vibert}, {Vielva}, {Villa},
  {Vittorio}, {Wandelt}, {Wehus}, {White}, {White}, {Zacchei}, \&
  {Zonca}}]{2018arXiv180706205P}
{Planck Collaboration}, {Akrami}, Y., {Arroja}, F., {et~al.}
  2018{\natexlab{b}}, arXiv e-prints, arXiv:1807.06205

\bibitem[{{Postman} {et~al.}(2012){Postman}, {Coe}, {Ben{\'{\i}}tez},
  {Bradley}, {Broadhurst}, {Donahue}, {Ford}, {Graur}, {Graves}, {Jouvel},
  {Koekemoer}, {Lemze}, {Medezinski}, {Molino}, {Moustakas}, {Ogaz}, {Riess},
  {Rodney}, {Rosati}, {Umetsu}, {Zheng}, {Zitrin}, {Bartelmann}, {Bouwens},
  {Czakon}, {Golwala}, {Host}, {Infante}, {Jha}, {Jimenez-Teja}, {Kelson},
  {Lahav}, {Lazkoz}, {Maoz}, {McCully}, {Melchior}, {Meneghetti}, {Merten},
  {Moustakas}, {Nonino}, {Patel}, {Reg{\"o}s}, {Sayers}, {Seitz}, \& {Van der
  Wel}}]{CLASH}
{Postman}, M., {Coe}, D., {Ben{\'{\i}}tez}, N., {et~al.} 2012, \apjs, 199, 25

\bibitem[{{Pratt} {et~al.}(2019){Pratt}, {Arnaud}, {Biviano}, {Eckert},
  {Ettori}, {Nagai}, {Okabe}, \& {Reiprich}}]{2019SSRv..215...25P}
{Pratt}, G.~W., {Arnaud}, M., {Biviano}, A., {et~al.} 2019, \ssr, 215, 25

\bibitem[{{Pratt} {et~al.}(2010){Pratt}, {Arnaud}, {Piffaretti},
  {B{\"o}hringer}, {Ponman}, {Croston}, {Voit}, {Borgani}, \& {Bower}}]{pra10}
{Pratt}, G.~W., {Arnaud}, M., {Piffaretti}, R., {et~al.} 2010, \aap, 511, A85

\bibitem[{{Pratt} {et~al.}(2016){Pratt}, {Pointecouteau}, {Arnaud}, \& {van der
  Burg}}]{2016A&A...590L...1P}
{Pratt}, G.~W., {Pointecouteau}, E., {Arnaud}, M., \& {van der Burg}, R.~F.~J.
  2016, \aap, 590, L1

\bibitem[{{Ruppin} {et~al.}(2017){Ruppin}, {Adam}, {Comis}, {Ade}, {Andr{\'e}},
  {Arnaud}, {Beelen}, {Beno{\^i}t}, {Bideaud}, {Billot}, {Bourrion}, {Calvo},
  {Catalano}, {Coiffard}, {D'Addabbo}, {De Petris}, {D{\'e}sert}, {Doyle},
  {Goupy}, {Kramer}, {Leclercq}, {Mac{\'{\i}}as-P{\'e}rez}, {Mauskopf},
  {Mayet}, {Monfardini}, {Pajot}, {Pascale}, {Perotto}, {Pisano},
  {Pointecouteau}, {Ponthieu}, {Pratt}, {Rev{\'e}ret}, {Ritacco}, {Rodriguez},
  {Romero}, {Schuster}, {Sievers}, {Triqueneaux}, {Tucker}, \& {Zylka}}]{rup17}
{Ruppin}, F., {Adam}, R., {Comis}, B., {et~al.} 2017, \aap, 597, A110

\bibitem[{{Ruppin} {et~al.}(2019{\natexlab{a}}){Ruppin}, {Mayet},
  {Mac{\'\i}as-P{\'e}rez}, \& {Perotto}}]{Ruppin2019a}
{Ruppin}, F., {Mayet}, F., {Mac{\'\i}as-P{\'e}rez}, J.~F., \& {Perotto}, L.
  2019{\natexlab{a}}, \mnras, 490, 784

\bibitem[{{Ruppin} {et~al.}(2018){Ruppin}, {Mayet}, {Pratt}, {Adam}, {Ade},
  {Andr{\'e}}, {Arnaud}, {Aussel}, {Bartalucci}, {Beelen}, {Beno{\^i}t},
  {Bideaud}, {Bourrion}, {Calvo}, {Catalano}, {Comis}, {De Petris},
  {D{\'e}sert}, {Doyle}, {Driessen}, {Goupy}, {Kramer}, {Lagache}, {Leclercq},
  {Lestrade}, {Mac{\'{\i}}as-P{\'e}rez}, {Mauskopf}, {Monfardini}, {Perotto},
  {Pisano}, {Pointecouteau}, {Ponthieu}, {Rev{\'e}ret}, {Ritacco}, {Romero},
  {Roussel}, {Schuster}, {Sievers}, {Tucker}, \& {Zylka}}]{rup18}
{Ruppin}, F., {Mayet}, F., {Pratt}, G.~W., {et~al.} 2018, \aap, 615, A112

\bibitem[{{Ruppin} {et~al.}(2019{\natexlab{b}}){Ruppin}, {Sembolini}, {De
  Petris}, {Adam}, {Cialone}, {Mac{\'\i}as-P{\'e}rez}, {Mayet}, {Perotto}, \&
  {Yepes}}]{2019A&A...631A..21R}
{Ruppin}, F., {Sembolini}, F., {De Petris}, M., {et~al.} 2019{\natexlab{b}},
  \aap, 631, A21

\bibitem[{{Salvati} {et~al.}(2020){Salvati}, {Douspis}, \&
  {Aghanim}}]{2020arXiv200510204S}
{Salvati}, L., {Douspis}, M., \& {Aghanim}, N. 2020, arXiv e-prints,
  arXiv:2005.10204

\bibitem[{{Salvati} {et~al.}(2019){Salvati}, {Douspis}, {Ritz}, {Aghanim}, \&
  {Babul}}]{2019A&A...626A..27S}
{Salvati}, L., {Douspis}, M., {Ritz}, A., {Aghanim}, N., \& {Babul}, A. 2019,
  \aap, 626, A27

\bibitem[{{Sarazin}(1988)}]{1988xrec.book.....S}
{Sarazin}, C.~L. 1988, {X-ray emission from clusters of galaxies}

\bibitem[{{Sembolini} {et~al.}(2013){Sembolini}, {Yepes}, {De Petris},
  {Gottl{\"o}ber}, {Lamagna}, \& {Comis}}]{2013MNRAS.429..323S}
{Sembolini}, F., {Yepes}, G., {De Petris}, M., {et~al.} 2013, \mnras, 429, 323

\bibitem[{{Sunyaev} \& {Zel'dovich}(1972)}]{SZ1972}
{Sunyaev}, R.~A. \& {Zel'dovich}, Y.~B. 1972, \apspr, 4, 173

\bibitem[{{Sunyaev} \& {Zel'dovich}(1980)}]{SZ1980}
{Sunyaev}, R.~A. \& {Zel'dovich}, Y.~B. 1980, \araa, 18, 537

\bibitem[{{Umetsu} {et~al.}(2014){Umetsu}, {Medezinski}, {Nonino}, {Merten},
  {Postman}, {Meneghetti}, {Donahue}, {Czakon}, {Molino}, {Seitz}, {Gruen},
  {Lemze}, {Balestra}, {Ben{\'{\i}}tez}, {Biviano}, {Broadhurst}, {Ford},
  {Grillo}, {Koekemoer}, {Melchior}, {Mercurio}, {Moustakas}, {Rosati}, \&
  {Zitrin}}]{Umetsu15}
{Umetsu}, K., {Medezinski}, E., {Nonino}, M., {et~al.} 2014, \apj, 795, 163

\bibitem[{{Umetsu} {et~al.}(2018){Umetsu}, {Sereno}, {Tam}, {Chiu}, {Fan},
  {Ettori}, {Gruen}, {Okumura}, {Medezinski}, {Donahue}, {Meneghetti}, {Frye},
  {Koekemoer}, {Broadhurst}, {Zitrin}, {Balestra}, {Ben{\'{\i}}tez}, {Higuchi},
  {Melchior}, {Mercurio}, {Merten}, {Molino}, {Nonino}, {Postman}, {Rosati},
  {Sayers}, \& {Seitz}}]{2018ApJ...860..104U}
{Umetsu}, K., {Sereno}, M., {Tam}, S.-I., {et~al.} 2018, \apj, 860, 104

\bibitem[{{Vikhlinin}(2006)}]{vik06b}
{Vikhlinin}, A. 2006, \apj, 640, 710

\bibitem[{{Vikhlinin} {et~al.}(2006){Vikhlinin}, {Kravtsov}, {Forman}, {Jones},
  {Markevitch}, {Murray}, \& {Van Speybroeck}}]{vik06}
{Vikhlinin}, A., {Kravtsov}, A., {Forman}, W., {et~al.} 2006, \apj, 640, 691

\bibitem[{{Zitrin} {et~al.}(2011){Zitrin}, {Broadhurst}, {Barkana}, {Rephaeli},
  \& {Ben{\'\i}tez}}]{Zitrin11}
{Zitrin}, A., {Broadhurst}, T., {Barkana}, R., {Rephaeli}, Y., \&
  {Ben{\'\i}tez}, N. 2011, \mnras, 410, 1939

\bibitem[{{Zitrin} {et~al.}(2009){Zitrin}, {Broadhurst}, {Umetsu}, {Coe},
  {Ben{\'{\i}}tez}, {Ascaso}, {Bradley}, {Ford}, {Jee}, {Medezinski},
  {Rephaeli}, \& {Zheng}}]{Zitrin09}
{Zitrin}, A., {Broadhurst}, T., {Umetsu}, K., {et~al.} 2009, \mnras, 396, 1985

\bibitem[{{Zitrin} {et~al.}(2015){Zitrin}, {Fabris}, {Merten}, {Melchior},
  {Meneghetti}, {Koekemoer}, {Coe}, {Maturi}, {Bartelmann}, {Postman},
  {Umetsu}, {Seidel}, {Sendra}, {Broadhurst}, {Balestra}, {Biviano}, {Grillo},
  {Mercurio}, {Nonino}, {Rosati}, {Bradley}, {Carrasco}, {Donahue}, {Ford},
  {Frye}, \& {Moustakas}}]{Zitrin15}
{Zitrin}, A., {Fabris}, A., {Merten}, J., {et~al.} 2015, \apj, 801, 44

\bibitem[{{Zitrin} {et~al.}(2013{\natexlab{a}}){Zitrin}, {Menanteau}, {Hughes},
  {Coe}, {Barrientos}, {Infante}, \& {Mandelbaum}}]{Zitrin13a}
{Zitrin}, A., {Menanteau}, F., {Hughes}, J.~P., {et~al.} 2013{\natexlab{a}},
  \apjl, 770, L15

\bibitem[{{Zitrin} {et~al.}(2013{\natexlab{b}}){Zitrin}, {Meneghetti},
  {Umetsu}, {Broadhurst}, {Bartelmann}, {Bouwens}, {Bradley}, {Carrasco},
  {Coe}, {Ford}, {Kelson}, {Koekemoer}, {Medezinski}, {Moustakas}, {Moustakas},
  {Nonino}, {Postman}, {Rosati}, {Seidel}, {Seitz}, {Sendra}, {Shu}, {Vega}, \&
  {Zheng}}]{Zitrin13}
{Zitrin}, A., {Meneghetti}, M., {Umetsu}, K., {et~al.} 2013{\natexlab{b}},
  \apjl, 762, L30

\end{thebibliography}

\end{document}